\documentclass[12pt]{article}

\usepackage{amstext,amsmath,amssymb,amsfonts,bbm,slashed}
\usepackage[latin1]{inputenc}
\usepackage{epsfig}
\usepackage{hyperref}
\usepackage{amsthm}
\usepackage{subfigure}
\usepackage{color}
\usepackage{multirow}

\usepackage{lscape}

\allowdisplaybreaks

\newtheorem{lemma}{Lemma}

\newtheorem{definition}{Definition}

\newtheorem{proposition}{Proposition}

\newcommand{\bea}{\begin{eqnarray}}
\newcommand{\eea}{\end{eqnarray}}
\newcommand{\beq}{\begin{equation}}
\newcommand{\eeq}{\end{equation}}

\newcommand{\bpro}{\begin{pro}}
	\newcommand{\epro}{\end{pro}}
\newcommand{\blem}{\begin{lem}}
	\newcommand{\elem}{\end{lem}}
\newcommand{\bdfn}{\begin{dfn}}
	\newcommand{\edfn}{\end{dfn}}
\newcommand{\bcor}{\begin{cor}}
	\newcommand{\ecor}{\end{cor}}
\newcommand{\bthm}{\begin{thm}}
	\newcommand{\ethm}{\end{thm}}
\newcommand{\bex}{\begin{ex}}
	\newcommand{\eex}{\end{ex}}
\newcommand{\brmk}{\begin{rmk}}
	\newcommand{\ermk}{\end{rmk}}
\newcommand{\bpr}{\begin{pr}}
	\newcommand{\epr}{\end{pr}}

\newcommand{\al}{\alpha}

\makeatletter

\begin{document}

\begin{center}
	
	{\LARGE\bf  Einstein field equation,  recursion operators, Noether and master symmetries in  conformable 	Poisson	manifolds}

	\vspace{15pt}
	
	{\large\bf\textit{Mahouton Norbert Hounkonnou$^{\bf a,b},$ Mahougnon Justin Landalidji$^{\bf a,b}$, and Melanija Mitrovi\'c$^{\bf a,b,c}$}
	}
	
	\vspace{15pt}
	
	$^{a}$ International Chair in Mathematical Physics and Applications (ICMPA-UNESCO Chair)\\ University of Abomey-Calavi,
	072 BP. 50, Cotonou, Benin Republic  \\
	\vspace{5pt}
$^{b}$ 	 International Center for Research and Advanced Studies in Mathematical and Computer Sciences and Applications (ICRASMCSA), 072 BP. 50 Cotonou, Benin Republic \\
	\vspace{5pt}
	$^{c}$ Center of Applied Mathematics of the Faculty of Mechanical Engineering (CAM-FMEN),
	University of Ni\v s,  Serbia
\\
\vspace{5pt}
	{ \it norbert.hounkonnou@cipma.uac.bj
	},
	{\it landalidjijustin@cipma.net
	}
{\it melanija.mitrovic@masfak.ni.ac.rs }
	\vspace{10pt}
\end{center}
\textbf{Abstract:} We show that a Minkowski phase space endowed with a bracket relatively to a conformable differential realizes a Poisson algebra, confering a bi-Hamiltonian structure to the resulting manifold. We infer that the related Hamiltonian vector field  is an infinitesimal Noether symmetry, and compute the corresponding deformed recursion operator. Besides, using  the Hamiltonian-Jacobi separability, we  construct recursion operators for  Hamiltonian vector fields in conformable Poisson-Schwarzschild and Friedmann-Lemaître-Robertson-Walker (FLRW) manifolds, and derive related constants of motion, Christoffel symbols, components of  Riemann and Ricci tensors, Ricci constant and components of Einstein tensor.
We highlight the existence of a hierarchy of bi-Hamiltonian structures in both the manifolds, and compute  a family of recursion operators and master symmetries generating the constants of motion.
\newline
\newline
\textbf{Keywords}: Einstein field equation,  recursion operator, Noether symmetry,  master symmetry, conformable differential,  Poisson manifold.
\newline
\newline
\textbf{ Mathematics Subject Classification (2010)}: 37C10; 37J35; 37K05;  37K10.

\section{Introduction}\label{sec1}
Conformable fractional calculus has a long and rich history.
In 1695, Gottfried Leibniz asked Guillaume l'H\^{o}spital if the (integer) order of derivatives and integrals could be extended \cite{val}. Was it possible if the order was some irrational, fractional or complex number? This idea motivated many mathematicians, physicists and engineers to develop the concept of fractional calculus in diverse fields of science and engineers, see $e.g.,$ \cite{chun, metz,herr, agra,alm,bale,iom,efe},  and
references therein. During four centuries, many famous mathematicians contributed to this development. It is still nowadays one of the most intensively developing areas of mathematical analysis, including several definitions
of fractional operators like Riemann-Liouville, Caputo, Gr\"{u}nwald-Letnikov, Riesz and Weyl definitions \cite{jah}, \cite{mac}, \cite{tar}, \cite{chun}.  Two  of these definitions, namely Riemann-Liouville and Caputo ones are famous. Mathematicians prefer Riemann-Liouville fractional derivative while physicists and engineers use Caputo fractional
one. Indeed, the Riemann-Liouville
fractional derivative of a constant is not zero, and it requires fractional initial conditions which are not generally specified \cite{chun}. In contrast, Caputo derivative of a constant is zero, and a fractional differential equation expressed in terms of Caputo fractional derivative requires standard boundary conditions. Unfortunately, the Riemann-Liouville derivative and Caputo derivative do not obey the Leibniz rule and chain rule, which sometimes prevent us from applying these derivatives to  ordinary physical systems with standard Newton derivative. In 2014, Khalil {\it et al.} \cite{kha} introduced the new fractional derivative called conformable fractional derivative and
integral obeying the Leibniz rule and chain rule. One year later, $i.e.,$ in 2015, Chung \cite{chun} used this conformable fractional derivative and integral to discuss the fractional version of the Newtonian
mechanics. In that work, he constructed the fractional Euler-Lagrange equation from the fractional version of the calculus of variations and used this equation to discuss some mechanical problems such as fractional harmonic oscillator problem, the fractional damped oscillator problem and the forced oscillator problem.
In 2017,  Chung {\it et al.} \cite{chun2} discussed the dynamics of a particle in a viscoelastic medium using the conformable fractional derivative of order $\alpha$ with respect to time. Further, in 2019, the same authors \cite{chun3} discussed the fractional classical mechanics and applied it to the anomalous diffusion relation from the $\alpha$-deformed Langevin equation. During the same year,  Kiskinov {\it et al.} \cite{kisk} investigated the Cauchy problem for nonlinear systems with conformable derivatives and variable delays. Furthermore, Khalil {\it et al.} gave the geometric meaning of conformable derivative via fractional cords in 2019 \cite{kha2}. In 2020, Chung {\it et al.} \cite{chun4} studied the deformed special relativity based on $\alpha$-deformed binary operations. In that work, they gave the $\alpha$-translation invariant
distance ($\alpha$-distance) of infinitesimally close space-time based on the definition of $\alpha$-translation invariant infinitesimal displacement and $\alpha$-translation invariant infinitesimal time interval. 

In addition, in the last few decades,  there was a renewed interest in
completely integrable Hamiltonian systems
(IHS), the concept of which goes back to Liouville in 1897 \cite{lio}
and Poincar\'{e} in 1899 \cite{pau}. In short, IHS are defined as nonlinear differential equations admitting a Hamiltonian description and possessing sufficiently many constants of motion so that they can be integrated by quadratures \cite{fil1}. This Liouville formalism  does not provide a method for obtaining the integrals of motion; therefore it has been
necessary to elaborate different methods for obtaining constants of motion (Hamilton-Jacobi
separability, Lax pairs formalism, Noether symmetries, Hidden symmetries, etc). 
A relevant progress in the analysis of the integrability was the important
remark that many of these systems are Hamiltonian dynamics with respect to two compatible symplectic structures  \cite{mag1,gel,vil1}, permitting a geometrical interpretation of the so-called recursion operator \cite{lax, hkn2, hkn3}.  A description of integrability
working both for systems with finitely many degrees of freedom and for field theory
can be given in terms of invariant, diagonalizable mixed $(1,1)$-tensor field, having bidimensional eigenspaces and vanishing Nijenhuis torsion. One of powerful methods of describing IHS with involutive Hamiltonian functions or constants of motion uses the recursion operator admitting a vanishing Nijenhuis torsion. In 2015,  Takeuchi  constructed  recursion operators of Hamiltonian vector fields of geodesic flows for some  Riemannian and Minkowski  metrics \cite{tak}, and  obtained related constants of motion. 
In his work, Takeuchi  used five particular solutions of the Einstein equation in the Schwarzschild,  Reissner-Nordstr\"{o}m,   Kerr,  Kerr-Newman, and  FLRW metrics, and  constructed recursion operators  inducing the complete integrability of the Hamiltonian functions. Further, in 2019, we  investigated  the same problem in a noncommutative Minkowski phase space \cite{hkn1}. 
  
In the present work, we investigate Noether symmetry and recursion operators induced by an $\alpha$-deformed Poisson algebra in a Minkowski phase space. We construct recursion operators using $\alpha-$deformed Schwarzschild and Friedmann-Lemaître-Robertson-Walker (FLRW) metrics and discuss their relevant master symmetries. 

The paper is organized as follows.  In Section \ref{sec2}, we
give the notion of conformable differential and related formulation of the wellknown
   Takeuchi Lemma \cite{tak}. In Section
\ref{sec3}, we construct an $\alpha$-deformed Poisson algebra and the Lie algebra of deformed vector fields, prove  the existence of infinitesimal Noether symmetry and  bi-Hamiltonian structure, and compute the corresponding recursion operator in an $\alpha$-deformed Minkowski phase space. In Section \ref{sec4}, we  construct recursion operators for   Hamiltonian vector fields, related constants of motion, Christoffel symbols, components of  Riemann and Ricci tensors, Ricci constant and components of Einstein tensor in the framework of  $\alpha$-deformed Schwarzschild and FLRW  metrics. In Section \ref{sec5},  we derive a hierarchy of master symmetries and compute the conserved quantities. In Section
\ref{sec6}, we end with some concluding remarks.
\section{Conformable differential and formulation of Takeuchi Lemma} \label{sec2}
	A Hamiltonian system is a triple $(\mathcal{Q}, \omega , H )$, where $(\mathcal{Q}, \omega)$ is a symplectic manifold and  $H$  is a smooth function
on $\mathcal{Q}$, called  {\it Hamiltonian} or  {\it Hamiltonian function} \cite{rud}.

Given a general dynamical system defined on the $2n$-dimensional manifold $\mathcal{Q}$ \cite{smir,smir2}, its evolution can be described by the equation
{\small\begin{equation} \label{dyn}
	\dot{x}(t) = X(x), \quad x \in \mathcal{Q}, \quad X \in \mathcal{T}\mathcal{Q}.
	\end{equation}}
If the system \eqref{dyn} admits two different Hamiltonian representations:
{\small\begin{equation*} \label{dyn2}
	\dot{x}(t) = X_{H_{1},H_{2}} = \mathcal{P}_{1}dH_{1} = \mathcal{P}_{2}dH_{2},
	\end{equation*}}
its integrability as well as many other properties are subject to Magri's approach. The bi-Hamiltonian vector field $X_{H_{1},H_{2}}$ is defined by two pairs of Poisson bivectors $\mathcal{P}_{1}, \mathcal{P}_{2}$ and Hamiltonian functions $ H_{1},H_{2}.$ Such a  manifold $\mathcal{Q}$ equipped with two Poisson bivectors is called a double Poisson manifold, and the quadruple $(\mathcal{Q},\mathcal{P}_{1},\mathcal{P}_{2},X_{H_{1},H_{2}} )$ is  called a bi-Hamiltonian system.   $\mathcal{P}_{1}$ and $\mathcal{P}_{2}$ are  two compatible Poisson bivectors with vanishing Schouten-Nijenhuis bracket \cite{du}: 
$$[\mathcal{P}_{1}, \mathcal{P}_{2}]_{NS} = 0.$$

A recursion operator 
$
	T : \mathcal{T}\mathcal{Q} \longrightarrow 	\mathcal{T}\mathcal{Q}
$
is defined by 
\begin{align*}
	T := \mathcal{P}_{2} \circ \mathcal{P}_{1}^{-1}.
\end{align*}

A Noether symmetry is a diffeomorphism 	$\Phi : \mathcal{Q} \longrightarrow  \mathcal{Q} $ such that \cite{rom}:
\begin{equation*}
 \  \Phi^{\ast} \omega = \omega,\quad 
\ \Phi^{\ast} H = H.
\end{equation*}
An infinitesimal Noether symmetry is a vector field 	$Y \in \mathfrak{X}( \mathcal{Q})$ (the set of all differentiable vector fields on $ \mathcal{Q}$) such that:
\begin{equation*}
 \  \mathcal{L}_{Y} \omega = 0, \quad 
 \	 \mathcal{L}_{Y} H = 0. 
\end{equation*}

\begin{definition}
 Consider the map $g$ and its inverse $ g^{-1} $: 
\begin{align*}
& g:  \mathbb{R}^{2n}_{\alpha}  \longrightarrow  \mathbb{R}^{2n}  \qquad \qquad \qquad \qquad g^{-1}:  \mathbb{R}^{2n}  \longrightarrow \mathbb{R}^{2n}_{\alpha} \\
& \ \ \qquad z \longmapsto g(z) = |z|^{\alpha - 1}z = \mathbf{Z} \qquad \qquad \ \mathbf{Z} \longmapsto g^{-1} (\mathbf{Z}) = |\mathbf{Z}|^{(1/\alpha) -1}\mathbf{Z} = z,
\end{align*}
where 
$g(0) = 0,\ g(1) = 1,$ and $ g(\pm \infty) = \pm \infty.$
Then, for this map, the $\alpha$-addition, $\alpha$-subtraction, $\alpha$-multiplication, and $\alpha$-division are given by:
\begin{align*}
& a \oplus_{\alpha} b = |a|a|^{\alpha - 1} + b|b|^{\alpha -1}|^{(1/\alpha) - 1} (a|a|^{\alpha - 1} + b|b|^{\alpha - 1}), \\
& a \ominus_{\alpha} b = |a|a|^{\alpha - 1} - b|b|^{\alpha -1}|^{(1/\alpha) - 1} (a|a|^{\alpha - 1} - b|b|^{\alpha - 1}),\\
& a \otimes_{\alpha} b = ab, \\
& a \oslash_{\alpha} b = \dfrac{a}{b},
\end{align*}
where $a,b \in \mathbb{R}^{2n}_{\alpha}$. 
\end{definition}
\begin{definition}
Let $h$ be a differentiable coordinates function on $\mathbb{R}^{2n}_{\alpha}$.  The conformable differential, also called $\alpha$-differential in sequel,  with respect to the position $q$ and its associated momentum $p$ is defined by:
\begin{align*}
d_{\alpha}: \mathbb{R}^{2n}_{\alpha}  &\longrightarrow  \mathbb{R}^{2n} \cr
 h & \longmapsto d_{\alpha}h:=  \sum_{\mu =1}^{2n}    \al |x_{\mu}|^{\alpha -1} \dfrac{\partial}{\partial x_{\mu}}h, \quad ( x_{\nu} = q^{\nu}, x_{\nu + n} = p_{\nu}, \ n = 4, \ \nu = 1,2,3,4)
\end{align*}
satisfying the following properties: 
\begin{itemize}
\item[(i)] $d_{\al} (ah + bf ) = ad_{\al}h + bd_{\al}f $ for all $ a,b \in \mathbb{R}$;
 \item[(ii)] $ d_{\al}(h^{m}) = m h^{m-1}  d_{\al}h,$ for all $m \in \mathbb{R}$;
 \item[(iii)] $ d_{\al}(c) = 0,$ for all constant functions $h(q,p) = c$;
 \item[(iv)] $ d_{\al}( hf) = hd_{\al}f + fd_{\al}h $ ;
 \item[(iv)] $ d_{\al}\bigg(\dfrac{h}{f}\bigg) = \dfrac{fd_{\al}h - hd_{\al}f  }{f^{2}}$,
 where $f$ is also a differentiable coordinates function on $\mathbb{R}^{2n}_{\alpha}$. 
\end{itemize}
\end{definition}

The $\alpha$-differential produces a new deformed phase space called  \textit{$\alpha$-deformed phase space.}
The ordinary differential is obtained for $\alpha = 1$.
Using the $\alpha$-addition and  $\alpha$-subtraction, we obtain the following infinitesimal distance between two points of coordinates $ (x_{i}, ...,x_{n})$ and $( x_{i}\oplus_{\alpha} d_{\alpha}x_{i}, ...., x_{n}\oplus_{\alpha} d_{\alpha}x_{n})$  
\begin{align*}\label{eq8}
d_{\alpha}s = (d^{2}_{\alpha}x_{i} + ....+d^{2}_{\alpha}x_{n})^{\frac{1}{2}}.
\end{align*} 

In the $\mathbb{R}_{\alpha}^{2n}$, Takeuchi Lemma \cite{tak} takes the following form:
\begin{lemma}\label{lem1}
	Consider $\alpha$-vector fields 
	\begin{equation*}
	X_{\alpha_{i}} = - |x_{i}|^{( 1- \alpha)}|x_{n +i}|^{(1 - \alpha )} \dfrac{\partial{}}{\partial{x_{n + i}}}, i = 1,...,n
	\end{equation*}
	on $\mathbb{R}_{\alpha}^{2n}$ and 
	\begin{equation*}
	T_{\alpha} = \sum_{i =1}^{n}|x_{i}|^{(  \alpha - 1)}|x_{i}\Bigg(\dfrac{\partial}{\partial{x_{i}}} \otimes dx_{i} + \dfrac{\partial}{\partial{x_{n+i}}} \otimes dx_{n+i}\Bigg),
	\end{equation*}
	a $(1,1)$-tensor field on $\mathbb{R}_{\alpha}^{2n}$.
	Then, we have that the Nijenhuis torsion of $T_{\alpha}$ is vanishing, $i.e.,$  $ \mathcal{N}_{T_{\alpha}} = 0$ and $\mathcal{L}_{X_{\alpha_{i}}} T_{\alpha} = 0,$
	that is, the $(1,1)$-tensor field $T_{\alpha}$ is an $\al$-recursion operator of $ X_{\alpha_{i}}, (i = 1,...,n)$.
\end{lemma}

\textbf{Proof.} 

We have:
\begin{align*}
\mathcal{L}_{X_{\alpha_{i}}}	T_{\alpha} & = \mathcal{L}_{X_{\alpha_{i}}}\Bigg\{\sum_{i =1}^{n}|x_{i}|^{(  \alpha - 1)}|x_{i}\bigg(\dfrac{\partial}{\partial{x_{i}}} \otimes dx_{i} + \dfrac{\partial}{\partial{x_{n+i}}} \otimes dx_{n+i}\bigg)\Bigg\} \\
& = \sum_{i =1}^{n}\Bigg\{\mathcal{L}_{X_{\alpha_{i}}}(|x_{i}|^{(  \alpha - 1)}|x_{i})\bigg(\dfrac{\partial}{\partial{x_{i}}} \otimes dx_{i} + \dfrac{\partial}{\partial{x_{n+i}}} \otimes dx_{n+i}\bigg) \\
&  + |x_{i}|^{(  \alpha - 1)}|x_{i}\bigg(\mathcal{L}_{X_{\alpha_{i}}}\bigg[\dfrac{\partial}{\partial{x_{i}}} \otimes dx_{i}\bigg] + \mathcal{L}_{X_{\alpha_{i}}}\bigg[\dfrac{\partial}{\partial{x_{n+i}}} \otimes dx_{n+i}\bigg]\bigg)\Bigg\} \\
\mathcal{L}_{X_{\alpha_{i}}}	T_{\alpha}& = \sum_{i =1}^{n}|x_{i}|^{(  \alpha - 1)}|x_{i}\bigg(\mathcal{L}_{X_{\alpha_{i}}}\bigg[\dfrac{\partial}{\partial{x_{i}}} \otimes dx_{i}\bigg] + \mathcal{L}_{X_{\alpha_{i}}}\bigg[\dfrac{\partial}{\partial{x_{n+i}}} \otimes dx_{n+i}\bigg]\bigg)\cr
\ \mbox{because} \ \mathcal{L}_{X_{\alpha_{i}}}(|x_{i}|^{(  \alpha - 1)}|x_{i})= 0.
\end{align*}
Then,
\small\begin{align*}
\mathcal{L}_{X_{\alpha_{i}}}T_{\alpha}&= \sum_{i =1}^{n}|x_{i}|^{(  \alpha - 1)}|x_{i}\Bigg(\mathcal{L}_{X_{\alpha_{i}}}\bigg[\dfrac{\partial}{\partial{x_{i}}}\bigg] \otimes dx_{i} +  \dfrac{\partial}{\partial{x_{i}}}\otimes \mathcal{L}_{X_{\alpha_{i}}}(dx_{i}) 
\cr & + \mathcal{L}_{X_{\alpha_{i}}}\bigg[\dfrac{\partial}{\partial{x_{n+i}}}\bigg] \otimes dx_{n+i} + \dfrac{\partial}{\partial{x_{n+i}}} \otimes \mathcal{L}_{X_{\alpha_{i}}}(dx_{n+i}) \Bigg)\cr
\mathcal{L}_{X_{\alpha_{i}}}T_{\alpha} & = 0.
\end{align*}

The components of the Nijenhuis torsion as follows \cite{tak}:
\begin{align*}
(\mathcal{N}_{T_{\alpha}})^{h}_{ij} & = (T_{\al})^{k}_{i}\dfrac{\partial{(T_{\al})^{h}_{j}}}{\partial{x_{k}}}  - (T_{\al})^{k}_{j}\dfrac{\partial{(T_{\al})^{h}_{i}}}{\partial{x_{k}}} + (T_{\al})^{h}_{k}\dfrac{\partial{(T_{\al})^{k}_{i}}}{\partial{x_{j}}}  - (T_{\al})^{h}_{k}\dfrac{\partial{(T_{\al})^{k}_{j}}}{\partial{x_{i}}} \\
& = |x_{i}|^{(  \alpha - 1)}|x_{i}\dfrac{\partial {(T_{\al})^{h}_{j}}}{\partial{x_{i}}} -  |x_{j}|^{(  \alpha - 1)}|x_{j}\dfrac{\partial {(T_{\al})^{h}_{i}}}{\partial{x_{j}}} +  (T_{\al})^{h}_{i}\dfrac{\partial {(|x_{i}|^{(  \alpha - 1)}|x_{i})}}{\partial{x_{j}}} -  (T_{\al})^{h}_{j}\dfrac{\partial {(|x_{j}|^{(  \alpha - 1)}|x_{j})}}{\partial{x_{i}}} \\
& = |x_{i}|^{(  \alpha - 1)}|x_{i}\dfrac{\partial {(T_{\al})^{h}_{j}}}{\partial{x_{i}}} -  |x_{j}|^{(  \alpha - 1)}|x_{j}\dfrac{\partial {(T_{\al})^{h}_{i}}}{\partial{x_{j}}} +  \al (T_{\al})^{h}_{i} |x_{i}|^{(  \alpha - 1)}\delta^{i}_{j} -  \al (T_{\al})^{h}_{j} |x_{j}|^{(  \alpha - 1)}\delta^{j}_{i}.
\end{align*}
\begin{enumerate}
	\item If $i=j,$ we have $\delta^{i}_{j} = \delta^{j}_{i} = 1$ and we get
	\begin{align} \label{Eq_2_5}
	(\mathcal{N}_{	T_{\alpha}})^{h}_{ij} & = |x_{i}|^{(  \alpha - 1)}|x_{i}\dfrac{\partial {(T_{\al})^{h}_{i}}}{\partial{x_{i}}} -  |x_{i}|^{(  \alpha - 1)}|x_{i}\dfrac{\partial {(T_{\al})^{h}_{i}}}{\partial{x_{i}}} +  \al |x_{i}|^{(  \alpha - 1)}(T_{\al})^{h}_{i} -  \al |x_{i}|^{(  \alpha - 1)}(T_{\al})^{h}_{i} = 0 ; 
	\end{align}
	\item If $i\neq j,$ we have $\delta^{i}_{j} = \delta^{j}_{i} = 0$ and
	$\dfrac{\partial {(T_{\al})^{h}_{j}}}{\partial{x_{i}}} = \dfrac{\partial {(T_{\al})^{h}_{i}}}{\partial{x_{j}}}= 0.$
	Then,
	\begin{equation}\label{Eq_2_6}
	(\mathcal{N}_{T_{\alpha}})^{h}_{ij} = 0.
	\end{equation}
\end{enumerate}
From \eqref{Eq_2_5} and \eqref{Eq_2_6},  we get $\mathcal{N}_{T_{\alpha}} = 0. $
$\hfill{\square}$
\section{Recursion operator in  $\al$-deformed Minkowski phase space} \label{sec3}
In this section, we derive the  recursion operator of Hamiltonian vector fields of geodesic flow for a free particle in an $\al$-deformed  Minkowski phase space and obtain the associated constants of motion.  
\subsection{Symplectic structure, Poisson bracket and  Lie algebra}
We  consider our configuration space as a manifold $\mathcal{Q} = \mathbb{R}_{\alpha}^{4} \backslash \{0\}$
that is, a four-dimensional real Euclidean vector space with the origin removed. 
The cotangent
bundle $\mathcal{T}^{\ast}\mathcal{Q}=\mathcal{Q} \times \mathbb{R}_{\alpha}^{4} $ has a  natural $\al$-symplectic structure  \\ $\omega_{\alpha} :
\mathcal{T}\mathcal{Q} \longrightarrow \mathcal{T}^{\ast}\mathcal{Q}$ which, in local coordinates  $(q,p)$, is given by
\begin{equation*}\label{Ksy}
\omega_{\alpha} = \sum_{\mu = 1}^{4} d_{\alpha}p_{\mu} \wedge  d_{\alpha}q^{\mu} = \sum_{\mu = 1}^{4} \al^{2}|p_{\mu}|^{\al - 1}|q^{\mu}|^{\al - 1}dp_{\mu} \wedge  dq^{\mu}.
\end{equation*}
Since $\omega_{\alpha}$ is
non-degenerate, it induces an  inverse map, called  $\al$-bivector field  $\mathcal{P}_{\alpha}$:
$\mathcal{T}^{\ast}\mathcal{Q} \longrightarrow \mathcal{T}\mathcal{Q}$ (tangent bundle)
defined by
\begin{equation*}\label{Kbi}
\mathcal{P}_{\alpha} = \sum_{\mu = 1}^{4} \al^{-2}|p_{\mu}|^{1 - \al }|q^{\mu}|^{1 -\al}\dfrac{\partial}{\partial p_{\mu}} \wedge \dfrac{\partial}{\partial q^{\mu}}
,\quad \omega_{\alpha} \circ \mathcal{P}_{\alpha} = \mathcal{P}_{\alpha} \circ \omega_{\alpha} = 1,
\end{equation*}
and used  to construct the  $\al$-Hamiltonian vector field $ X_{\alpha _{f}}$
of an $\al$-Hamiltonian function $f$ by the relation
\begin{equation*}\label{Kvec}
X_{\alpha _{f}} = \mathcal{P}_{\alpha}df. 
\end{equation*} 

We consider now  the next $\al$-Minkowski metric on the manifold  $ \mathcal{Q}$: 
\begin{equation*}
d_{\alpha} s^{2} = -  \alpha^{2}|q^{1}|^{2(\alpha - 1)}(dq^{1})^{2} +  \alpha^{2}|q^{2}|^{2(\alpha - 1)}(dq^{2})^{2} +  \alpha^{2}|q^{3}|^{2(\alpha - 1)}(dq^{3})^{2}+  \alpha^{2}|q^{4}|^{2(\alpha - 1)}(dq^{4})^{2},
\end{equation*}
where $c = 1$ for commodity
yielding
the tensor metric $ (g_{\mu\nu})_{\alpha}$  and its inverse $(g^{\mu\nu})_{\alpha}$
\begin{equation*}
(g_{\mu\nu})_{\alpha} = \alpha^{2}\left(
\begin{array}{cccc}
- (q^{1})^{2(\alpha - 1)} & 0 & 0 & 0 \\
0 & (q^{2})^{2(\alpha - 1)}& 0 & 0 \\
0 & 0 & (q^{3})^{2(\alpha - 1)} & 0 \\
0 & 0 & 0 & (q^{4})^{2(\alpha - 1)} \\
\end{array}
\right),
\end{equation*}

\begin{equation*}
(g^{\mu\nu})_{\alpha} = \dfrac{1}{\alpha^{2}}\left(
\begin{array}{cccc}
- (q^{1})^{2( 1 - \alpha )} & 0 & 0 & 0 \\
0 & (q^{2})^{2( 1 - \alpha )}& 0 & 0 \\
0 & 0 & (q^{3})^{2( 1 - \alpha )} & 0 \\
0 & 0 & 0 & (q^{4})^{2( 1 - \alpha )} \\
\end{array}
\right).
\end{equation*}
In our framework, the equation of the geodesic on the manifold $\mathcal{Q}$ is given by
\begin{equation} \label{geod}
\dfrac{d^{2}q^{\mu}}{dt^{2}} + (\Gamma^{\mu}_{\nu\lambda})_{\alpha}\dfrac{dq^{\nu}}{dt}\dfrac{dq^{\lambda}}{dt} = 0, \ \ (\nu,\mu, \lambda = 1,2,3,4),
\end{equation}
 where
\begin{equation}\label{Chris}
(\Gamma^{\mu}_{\nu\lambda})_{\alpha} = \dfrac{1}{2}(g^{\mu \epsilon})_{\alpha}\Bigg(\dfrac{\partial{(g_{\epsilon\nu})_{\alpha}}}{\partial{q^{\lambda}}} + \dfrac{\partial{(g_{\epsilon \lambda})_{\alpha}}}{\partial{q^{\nu}}} - \dfrac{\partial{(g_{\nu \lambda})_{\alpha}}}{\partial{q^{\epsilon}}} \Bigg)
\end{equation}
are $\al$-Christoffel symbols. From \eqref{Chris}, we have
{\small\begin{equation*}
	(\Gamma^{1}_{11})_{\alpha} = \dfrac{\alpha - 1}{q^{1}}; \
	(\Gamma^{2}_{22})_{\alpha} = \dfrac{\alpha - 1}{q^{2}};\
	(\Gamma^{3}_{33})_{\alpha} = \dfrac{\alpha - 1}{q^{3}};\
	(\Gamma^{4}_{44})_{\alpha} = \dfrac{\alpha - 1}{q^{4}}; \
	(\Gamma^{\mu}_{\nu\lambda})_{\alpha} = 0, \ \ \mbox{otherwise},
	\end{equation*} }
and obtain that the $\al$-Riemann tensor components are vanished, $i.e.,  R_{ijkl} = 0, \ (i,j,k,l = 1,2,3,4 )$. Then,  the Minkowski phase space endowed with the metric $d_{\alpha} s^{2} $ is a flat space. Thus, we notice that this $\alpha$-deformation does not change the geometric structure of the ordinary Minkowski phase space. Further, the presence of the $\al$-Christoffel symbols $(\Gamma^{i}_{ii})_{\alpha}, \ (i = 1,2,3,4)$ means that the parallel displacement of any basic vector of our considered manifold with respect to itself always remains parallel with this same basic vector. The ordinary Minkowski phase space is obtained for $\al = 1.$

Since the quantities $ (\tilde{\Gamma}^{\mu}_{\nu\lambda})_{\alpha} =  \dfrac{1}{\alpha + 1}(\Gamma^{\mu}_{\nu\lambda})_{\alpha}  $ do not change  the geometric structure of the Minkowski phase space, we replace $(\Gamma^{\mu}_{\nu\lambda})_{\alpha} $ by $(\tilde{\Gamma}^{\mu}_{\nu\lambda})_{\alpha}$ in \eqref{geod}. Then, the equation of the geodesic becomes: 

\begin{equation*} \label{geod2}
\dfrac{d^{2}q^{\mu}}{dt^{2}} + (\tilde{\Gamma}^{\mu}_{\nu\lambda})_{\alpha}\dfrac{dq^{\nu}}{dt}\dfrac{dq^{\lambda}}{dt} = 0, \ \ (\nu,\mu, \lambda = 1,2,3,4).
\end{equation*}

If we put $\upsilon^{\mu} = \dfrac{dq^{\mu}}{dt}$,  we have a first order differential equation on the tangent bundle $\mathcal{T}(\mathcal{Q})$ of the manifold $\mathcal{Q}$:
{\small\begin{equation*}
	\dot{q^{\mu}} = \upsilon^{\mu}, \quad \dot{\upsilon^{\mu}} = - \dfrac{1}{\alpha + 1}(\Gamma^{\mu}_{\nu\lambda})_{\alpha}\upsilon^{\nu}\upsilon^{\lambda} .
	\end{equation*}}
From the above equations, we get the $\al$-geodesic spray
{\small\begin{equation*}
	X_{\alpha} := \upsilon^{\mu}\dfrac{\partial}{\partial{q^{\mu}}} -\dfrac{1}{\alpha + 1}(\Gamma^{\mu}_{\nu\lambda})_{\alpha}\upsilon^{\nu}\upsilon^{\lambda}\dfrac{\partial}{\partial{\upsilon^{\mu}}}.
	\end{equation*}}
By setting $p_{\mu} = \varepsilon_{\mu\epsilon}\upsilon^{\epsilon},$ $\varepsilon = sgn(-,+,+,+),$ the $\al$-vector field $X_{\alpha}$ is equivalently transformed to the $\al$-vector field $X_{\alpha}$ on the cotangent bundle $\mathcal{T}^{\ast}(\mathcal{Q})$ such that
{\small\begin{equation}\label{Hvec1}
	X_{\alpha} = - p_{1}\dfrac{\partial}{\partial{q^{1}}} + \sum^{4}_{k = 2}p_{k}\dfrac{\partial}{\partial{q^{k}}} + \bigg(\dfrac{\alpha - 1}{\alpha + 1}\bigg) \dfrac{p_{1}^{2}}{q^{1}} \dfrac{\partial}{\partial{p_{1}}} -  \sum^{4}_{k = 2} \bigg(\dfrac{\alpha - 1}{\alpha + 1}\bigg) \dfrac{p_{k}^{2}}{q^{k}} \dfrac{\partial}{\partial{p_{k}}},
	\end{equation}}
The vector field $X_{\alpha}$ is an $\al$-Hamiltonian vector field of a certain $\al$-Hamiltonian function $H_{\alpha}$.  
\begin{proposition}\label{prop1}
	The set 	$ \mathfrak{F}$ of differentiable functions defined on 	$\mathcal{T}^{\ast}(\mathcal{Q})$ endowed with the bracket 
	\begin{equation*} \label{conPoi}
	\{f,g\}_{\alpha}:= \sum_{\mu = 1}^{4}\alpha^{-2}|p_{\mu}|^{( 1- \alpha)}|q^{\mu}|^{(1 - \alpha )} \Bigg(\dfrac{\partial{f}}{\partial{p_{\mu}}}\dfrac{\partial{g}}{\partial{q^{\mu}}} - \dfrac{\partial{f}}{\partial{q^{\mu}}}\dfrac{\partial{g}}{\partial{p_{\mu}}}\Bigg)
	\end{equation*}
	is an  $\al$-deformed Poisson algebra.
\end{proposition}
\textbf{Proof.}	\\
To prove this Proposition, we just have to prove that the bracket 	$ 	\{.,.\}_{\alpha} $ is an $\al$-deformed Poisson bracket.

Let us consider  $f,g,$ and $h$ be three arbitrary elements of $ \mathfrak{F}$.
\begin{itemize}
  \item\textbf{Antisymmetry}
  {\small\begin{align*}
  \{f,g\}_{\alpha} &= \sum_{\mu = 1}^{4}\alpha^{-2}|p_{\mu}|^{( 1- \alpha)}|q^{\mu}|^{(1 - \alpha )} \Bigg(\dfrac{\partial{f}}{\partial{p_{\mu}}}\dfrac{\partial{g}}{\partial{q^{\mu}}} - \dfrac{\partial{f}}{\partial{q^{\mu}}}\dfrac{\partial{g}}{\partial{p_{\mu}}}\Bigg), \cr 
   & = - \sum_{\mu = 1}^{4}\alpha^{-2}|p_{\mu}|^{( 1- \alpha)}|q^{\mu}|^{(1 - \alpha )} \Bigg(\dfrac{\partial{g}}{\partial{p_{\mu}}}\dfrac{\partial{f}}{\partial{q^{\mu}}} - \dfrac{\partial{g}}{\partial{q^{\mu}}}\dfrac{\partial{f}}{\partial{p_{\mu}}}\Bigg), \cr
    & = -   \{g,f\}_{\alpha}.
     \end{align*}} 
\item\textbf{ Jacobi identity}
 {\small\begin{align}\label{ja1}
\{f,\{g, h \}_{\alpha}\}_{\alpha} & = \bigg\{ f, \sum_{\mu = 1}^{4}\alpha^{-2}|p_{\mu}|^{( 1- \alpha)}|q^{\mu}|^{(1 - \alpha )} \Bigg(\dfrac{\partial{g}}{\partial{p_{\mu}}}\dfrac{\partial{h}}{\partial{q^{\mu}}} - \dfrac{\partial{g}}{\partial{q^{\mu}}}\dfrac{\partial{h}}{\partial{p_{\mu}}}\Bigg) \bigg\}_{\alpha} \cr
& = \sum_{\mu,\nu = 1}^{4}\alpha^{-4}|p_{\nu}|^{( 1- \alpha)}|q^{\nu}|^{(1 - \alpha )}\Bigg[\dfrac{\partial{f}}{\partial{p_{\nu}}}\Bigg(\sigma_{1}|p_{\mu}|^{( 1- \alpha)}(q^{\mu})^{-\alpha}\bigg( \dfrac{\partial{g}}{\partial{p_{\mu}}}\dfrac{\partial{h}}{\partial{q^{\mu}}} - \dfrac{\partial{g}}{\partial{q^{\mu}}}\dfrac{\partial{h}}{\partial{p_{\mu}}} \bigg) \cr
& + |p_{\mu}|^{( 1- \alpha)}|q^{\mu}|^{(1 - \alpha )} \bigg( \dfrac{\partial^{2} g}{\partial q^{\nu} \partial p_{\mu}} \dfrac{\partial h}{\partial q^{\mu}} + \dfrac{\partial g}{ \partial p_{\mu}} \dfrac{\partial^{2} h}{\partial q^{\nu} \partial q^{\mu}} - \dfrac{\partial^{2} g}{\partial q^{\nu} \partial q^{\mu}}\dfrac{\partial h}{ \partial p_{\mu}} -\dfrac{\partial g}{\partial q^{\mu}}\dfrac{\partial^{2} h}{\partial q^{\nu} \partial p_{\mu}} \bigg)\Bigg)   \cr
& - \dfrac{\partial f}{\partial q^{\nu}} \Bigg(\sigma_{2}|q^{\mu}|^{( 1- \alpha)} (p_{\mu})^{-\alpha}  \bigg( \dfrac{\partial{g}}{\partial{p_{\mu}}}\dfrac{\partial{h}}{\partial{q^{\mu}}} - \dfrac{\partial{g}}{\partial{q^{\mu}}}\dfrac{\partial{h}}{\partial{p_{\mu}}} \bigg)  + |p_{\mu}|^{( 1- \alpha)}|q^{\mu}|^{(1 - \alpha )}  \cr
&\times \bigg( \dfrac{\partial^{2} g}{ \partial p_{\nu}\partial p_{\mu}} \dfrac{\partial h}{\partial q^{\mu}} + \dfrac{\partial g}{ \partial p_{\mu}} \dfrac{\partial^{2} h}{\partial p_{\nu} \partial q^{\mu}} - \dfrac{\partial^{2} g}{\partial p_{\nu} \partial q^{\mu}}\dfrac{\partial h}{ \partial p_{\mu}} -\dfrac{\partial g}{\partial q^{\mu}}\dfrac{\partial^{2} h}{\partial p_{\nu} \partial p_{\mu}} \bigg)\Bigg)  \Bigg], 
  \end{align} } 
  {\small\begin{align}\label{ja2}
 \{h,\{f, g \}_{\alpha}\}_{\alpha} & = \bigg\{ h, \sum_{\mu = 1}^{4}\alpha^{-2}|p_{\mu}|^{( 1- \alpha)}|q^{\mu}|^{(1 - \alpha )} \Bigg(\dfrac{\partial{f}}{\partial{p_{\mu}}}\dfrac{\partial{g}}{\partial{q^{\mu}}} - \dfrac{\partial{f}}{\partial{q^{\mu}}}\dfrac{\partial{g}}{\partial{p_{\mu}}}\Bigg) \bigg\}_{\alpha} \cr
                    & = \sum_{\mu,\nu = 1}^{4}\alpha^{-4}|p_{\nu}|^{( 1- \alpha)}|q^{\nu}|^{(1 - \alpha )}\Bigg[\dfrac{\partial{h}}{\partial{p_{\nu}}}\Bigg(\sigma_{1}|p_{\mu}|^{( 1- \alpha)} (q^{\mu})^{-\alpha}\bigg( \dfrac{\partial{f}}{\partial{p_{\mu}}}\dfrac{\partial{g}}{\partial{q^{\mu}}} - \dfrac{\partial{f}}{\partial{q^{\mu}}}\dfrac{\partial{g}}{\partial{p_{\mu}}} \bigg) \cr
                    & + |p_{\mu}|^{( 1- \alpha)}|q^{\mu}|^{(1 - \alpha )} \bigg( \dfrac{\partial^{2} f}{\partial q^{\nu} \partial p_{\mu}} \dfrac{\partial g}{\partial q^{\mu}} + \dfrac{\partial f}{ \partial p_{\mu}} \dfrac{\partial^{2} g}{\partial q^{\nu} \partial q^{\mu}} - \dfrac{\partial^{2} f}{\partial q^{\nu} \partial q^{\mu}}\dfrac{\partial g}{ \partial p_{\mu}} -\dfrac{\partial f}{\partial q^{\mu}}\dfrac{\partial^{2} g}{\partial q^{\nu} \partial p_{\mu}} \bigg)\Bigg)   \cr
                    & - \dfrac{\partial h}{\partial q^{\nu}} \Bigg( \sigma_{2}|q^{\mu}|^{( 1- \alpha)} (p_{\mu})^{-\alpha}  \bigg( \dfrac{\partial{f}}{\partial{p_{\mu}}}\dfrac{\partial{g}}{\partial{q^{\mu}}} - \dfrac{\partial{f}}{\partial{q^{\mu}}}\dfrac{\partial{g}}{\partial{p_{\mu}}} \bigg)  + |p_{\mu}|^{( 1- \alpha)}|q^{\mu}|^{(1 - \alpha )}  \cr
                 & \times\bigg( \dfrac{\partial^{2} f}{ \partial p_{\nu}\partial p_{\mu}} \dfrac{\partial g}{\partial q^{\mu}} + \dfrac{\partial f}{ \partial p_{\mu}} \dfrac{\partial^{2} g}{\partial p_{\nu} \partial q^{\mu}} - \dfrac{\partial^{2} f}{\partial p_{\nu} \partial q^{\mu}}\dfrac{\partial g}{ \partial p_{\mu}} -\dfrac{\partial f}{\partial q^{\mu}}\dfrac{\partial^{2} g}{\partial p_{\nu} \partial p_{\mu}} \bigg)\Bigg)  \Bigg], 
  \end{align} } 
      
 {\small\begin{align}\label{ja3}
 \{g,\{h, f \}_{\alpha}\}_{\alpha} & = \bigg\{ g, \sum_{\mu = 1}^{4}\alpha^{-2}|p_{\mu}|^{( 1- \alpha)}|q^{\mu}|^{(1 - \alpha )} \Bigg(\dfrac{\partial{h}}{\partial{p_{\mu}}}\dfrac{\partial{f}}{\partial{q^{\mu}}} - \dfrac{\partial{h}}{\partial{q^{\mu}}}\dfrac{\partial{f}}{\partial{p_{\mu}}}\Bigg) \bigg\}_{\alpha} \cr
                    & = \sum_{\mu,\nu = 1}^{4}\alpha^{-4}|p_{\nu}|^{( 1- \alpha)}|q^{\nu}|^{(1 - \alpha )}\Bigg[\dfrac{\partial{g}}{\partial{p_{\nu}}}\Bigg( \sigma_{1}|p_{\mu}|^{( 1- \alpha)} (q^{\mu})^{-\alpha}\bigg( \dfrac{\partial{h}}{\partial{p_{\mu}}}\dfrac{\partial{f}}{\partial{q^{\mu}}} - \dfrac{\partial{h}}{\partial{q^{\mu}}}\dfrac{\partial{f}}{\partial{p_{\mu}}} \bigg) \cr
                    & + |p_{\mu}|^{( 1- \alpha)}|q^{\mu}|^{(1 - \alpha )} \bigg( \dfrac{\partial^{2} h}{\partial q^{\nu} \partial p_{\mu}} \dfrac{\partial f}{\partial q^{\mu}} + \dfrac{\partial h}{ \partial p_{\mu}} \dfrac{\partial^{2} f}{\partial q^{\nu} \partial q^{\mu}} - \dfrac{\partial^{2} h}{\partial q^{\nu} \partial q^{\mu}}\dfrac{\partial f}{ \partial p_{\mu}} -\dfrac{\partial h}{\partial q^{\mu}}\dfrac{\partial^{2} f}{\partial q^{\nu} \partial p_{\mu}} \bigg)\Bigg)   \cr
                    & - \dfrac{\partial g}{\partial q^{\nu}} \Bigg( \sigma_{2}|q^{\mu}|^{( 1- \alpha)} (p_{\mu})^{-\alpha}  \bigg( \dfrac{\partial{h}}{\partial{p_{\mu}}}\dfrac{\partial{f}}{\partial{q^{\mu}}} - \dfrac{\partial{h}}{\partial{q^{\mu}}}\dfrac{\partial{f}}{\partial{p_{\mu}}} \bigg) 
                  + |p_{\mu}|^{( 1- \alpha)}|q^{\mu}|^{(1 - \alpha )}  \cr &\times\bigg( \dfrac{\partial^{2} h}{ \partial p_{\nu}\partial p_{\mu}} \dfrac{\partial f}{\partial q^{\mu}} + \dfrac{\partial h}{ \partial p_{\mu}} \dfrac{\partial^{2} f}{\partial p_{\nu} \partial q^{\mu}} - \dfrac{\partial^{2} h}{\partial p_{\nu} \partial q^{\mu}}\dfrac{\partial f}{ \partial p_{\mu}} -\dfrac{\partial h}{\partial q^{\mu}}\dfrac{\partial^{2} f}{\partial p_{\nu} \partial p_{\mu}} \bigg)\Bigg)  \Bigg],
  \end{align} } 
where $ \sigma_{1} = (1 - \al)(sgn(q^{\mu}))^{1 -\alpha}$ and $ \sigma_{2} = (1 - \al)(sgn(p_{\mu}))^{1 -\alpha}.$ 

 Summing \eqref{ja1}, \eqref{ja2}, and \eqref{ja3}, we get  
     
  \begin{equation*}
  \{f,\{g, h \}_{\alpha} \}_{\alpha} + \{g,\{h, f \}_{\alpha} \}_{\alpha} + \{h,\{f, g \}_{\alpha} \}_{\alpha} = 0,
  \end{equation*}  
  which is the Jacobi identity. 
   \item\textbf{ Derivation}
   \begin{align*}
   \{f,gh \}_{\alpha} & = \sum_{\mu = 1}^{4}\alpha^{-2}|p_{\mu}|^{( 1- \alpha)}|q^{\mu}|^{(1 - \alpha )} \Bigg(\dfrac{\partial{f}}{\partial{p_{\mu}}}\dfrac{\partial{(gh)}}{\partial{q^{\mu}}} - \dfrac{\partial{f}}{\partial{q^{\mu}}}\dfrac{\partial{(gh)}}{\partial{p_{\mu}}}\Bigg) \\
    & = \sum_{\mu = 1}^{4}\alpha^{-2}|p_{\mu}|^{( 1- \alpha)}|q^{\mu}|^{(1 - \alpha )}\Bigg[\dfrac{\partial{f}}{\partial{p_{\mu}}}\bigg(\dfrac{\partial{g}}{\partial{q^{\mu}}}h + g\dfrac{\partial{h}}{\partial{q^{\mu}}}\bigg) - \dfrac{\partial{f}}{\partial{q^{\mu}}}\bigg(\dfrac{\partial{g}}{\partial{p_{\mu}}}h + g\dfrac{\partial{h}}{\partial{p_{\mu}}}\bigg)\Bigg] \\
    & = \sum_{\mu = 1}^{4}\alpha^{-2}|p_{\mu}|^{( 1- \alpha)}|q^{\mu}|^{(1 - \alpha )}\Bigg[g\bigg(\dfrac{\partial{f}}{\partial{p_{\mu}}}\dfrac{\partial{h}}{\partial{q^{\mu}}} - \dfrac{\partial{f}}{\partial{q^{\mu}}}\dfrac{\partial{h}}{\partial{p_{\mu}}}\bigg) + \bigg(\dfrac{\partial{f}}{\partial{p_{\mu}}}\dfrac{\partial{g}}{\partial{q^{\mu}}} - \dfrac{\partial{f}}{\partial{q^{\mu}}}\dfrac{\partial{g}}{\partial{p_{\mu}}}\bigg)h\Bigg],
   \end{align*}
   which proves the derivative property: $ \{f,gh \}_{\alpha} = g\{f,h \}_{\alpha} + \{f,g \}_{\alpha}h.$
 
\end{itemize} 

Thus, the bracket 	$ 	\{.,.\}_{\alpha} $ is antisymmetry and satisfies the Jacobi identity and the derivation property. Therefore, it  is a Poisson bracket and $(\mathfrak{F}, \{.,.\}_{\alpha})$ is an $\al$-deformed Poisson algebra.
$\hfill{\square}$ 
\begin{proposition} \label{prop2}
	The set of $\al$-Hamiltonian vector fields 	$\mathfrak{X}_{\alpha_{\mathfrak{F}}}$ endowed with the Lie bracket given by the commutator $[.,.]$	 is an $\al$-deformed Lie algebra.
\end{proposition}
\textbf{Proof.}	\\
Using the Jacoby identity, we have:
\begin{align*}
\{f,\{g, h \}_{\alpha} \}_{\alpha} + \{g,\{h, f \}_{\alpha} \}_{\alpha} + \{h,\{f, g \}_{\alpha} \}_{\alpha} = 0.
\end{align*} 
The left hand side of this identity can be handled as:
\begin{align*}
\{f,\{g, h \}_{\alpha} \}_{\alpha} + \{g,\{h, f \}_{\alpha} \}_{\alpha} + \{h,\{f, g \}_{\alpha} \}_{\alpha} 
& = \{f,\{g, h \}_{\alpha} \}_{\alpha} - \{g,\{f, h \}_{\alpha} \}_{\alpha} - \{\{f, g \}_{\alpha},h\}_{\alpha} \cr
& = X_{\alpha _{f}}\{g, h \}_{\alpha} - \{g,  X_{\alpha _{f}}h \}_{\alpha} -  \{X_{\alpha _{f}}g,h\}_{\alpha}\cr
& = X_{\alpha _{f}}X_{\alpha _{g}}h - X_{\alpha _{g}}X_{\alpha _{f}}h - X_{\alpha _{\{f, g \}_{\alpha}}}h \cr
& = [X_{\alpha _{f}}, X_{\alpha _{g}}]h - X_{\alpha _{\{f, g \}_{\alpha}}}h
\end{align*} 
leading to 
\begin{align*}
[X_{\alpha _{f}}, X_{\alpha _{g}}]h = X_{\alpha _{\{f, g \}_{\alpha}}}h.
\end{align*}
Then, the map 
$f \mapsto  X_{\alpha _{f}} = \{f, . \}_{\alpha}, \quad \{f, . g \}_{\alpha} \mapsto  X_{\alpha _{\{f, g \}_{\alpha}}}$ 
is an $\al$-deformed Lie algebra morphism \\	$(\mathfrak{F}, \{.,.\}_{\alpha}) \rightarrow (\mathfrak{X}_{\alpha_{\mathfrak{F}}}, [.,.])$.
Therefore, 	$(\mathfrak{X}_{\alpha_{\mathfrak{F}}}, [.,.])$ is an $\al$-deformed Lie algebra.
$\hfill{\square}$  
\subsection{Noether symmetry and recursion operator} 
By definition, we have
\begin{equation} \label{Hvec2}
X_{\alpha}:=  \{H_{\alpha}, .\}_{\alpha} = \sum_{\mu = 1}^{4}\alpha^{-2}|p_{\mu}|^{( 1- \alpha)}|q^{\mu}|^{(1 - \alpha )} \Bigg(\dfrac{\partial{H_{\alpha}}}{\partial{p_{\mu}}}\dfrac{\partial}{\partial{q^{\mu}}} - \dfrac{\partial{H_{\alpha}}}{\partial{q^{\mu}}}\dfrac{\partial}{\partial{p_{\mu}}}\Bigg).
\end{equation}
Using \eqref{Hvec1} and \eqref{Hvec2}, we obtain the following set of equations:
{\small\begin{equation*} \label{sys1}
	\begin{cases}
	\alpha^{-2}|p_{1}|^{( 1- \alpha)}|q^{1}|^{(1 - \alpha )}\dfrac{\partial{H_{\alpha}}}{\partial{p_{1}}}  &= -p_{1} \\
	\alpha^{-2}|p_{1}|^{( 1- \alpha)}|q^{1}|^{(1 - \alpha )}\dfrac{\partial{H_{\alpha}}}{\partial{q^{1}}} &= - \bigg(\dfrac{\alpha - 1}{\alpha + 1}\bigg) \dfrac{p_{1}^{2}}{q^{1}}\\
	\alpha^{-2}|p_{k}|^{( 1- \alpha)}|q^{k}|^{(1 - \alpha )}\dfrac{\partial{H_{\alpha}}}{\partial{p_{k}}} &= p_{k},   \qquad \qquad \qquad k = 2,3,4  \\
	\alpha^{-2}|p_{k}|^{( 1- \alpha)}|q^{k}|^{(1 - \alpha )}\dfrac{\partial{H_{\alpha}}}{\partial{q^{k}}} &=  \bigg(\dfrac{\alpha - 1}{\alpha + 1}\bigg) \dfrac{p_{k}^{2}}{q^{k}}, \qquad  k = 2,3,4
	\end{cases}  
	\end{equation*} }
leading to 
{\small\begin{equation*}  
	H_{\alpha} = - \dfrac{ \alpha^{2}}{\alpha + 1} |q^{1}|^{\alpha -1}|p_{1}|^{\alpha + 1} + \sum_{k = 2}^{4} \dfrac{\alpha^{2}}{\alpha + 1} |q^{k}|^{\alpha -1}|p_{k}|^{\alpha + 1}.
	\end{equation*}}
This function is called the $\al$-Hamiltonian function. For $\alpha = 1$, we naturally obtain the Hamiltonian function of a free particle on the ordinary Minkowski phase space.

The vector field
{\small \begin{equation*}
	Y_{\alpha} =  - \dfrac{1}{2\alpha^{2}} |p_{1}|^{1 - \alpha} p_{1}^{-2} |q^{1}|^{1-\alpha} |q^{1}|^{\frac{1 - \alpha}{ 1 + \alpha}} \dfrac{\partial}{\partial p_{1}} + \dfrac{1}{2\alpha^{2}}\sum_{k=2}^{4} |p_{k}|^{1 - \alpha} p_{k}^{-2} |q^{k}|^{1-\alpha} |q^{k}|^{\frac{1 - \alpha}{ 1 + \alpha}} \dfrac{\partial}{\partial p_{k}}
	\end{equation*}}
is a master symmetry, $i.e.,$
$$ \left[\left[Y_{\alpha},X_{\alpha}\right],X_{\alpha} \right]=0,$$ 
and  the following  relations hold:
{\small \begin{align*}
	L_{\alpha}&:= \mathcal{L}_{Y_{\alpha}} H_{\alpha}  = \dfrac{1}{2}\bigg( p_{1}^{-1}(q^{1})^{\frac{1 - \alpha}{1 + \alpha}} + \sum_{k=2}^{4}  p_{k}^{-1}(q^{k})^{\frac{1 - \alpha}{1 + \alpha}} \bigg), \\
	\omega_{\alpha_{1}}& := \mathcal{L}_{Y_{\alpha}} \omega_{\alpha}=  d\iota_{Y_{\alpha}}\omega_{\alpha} +  \iota_{Y_{\alpha}}d\omega_{\alpha}  \cr
	& = p_{1}^{-3}|q^{1}|^{\frac{1 - \alpha}{1 + \alpha}} dp_{1} \wedge dq^{1} - \sum_{k = 2}^{4} p_{k}^{-3}|q^{k}|^{\frac{1 - \alpha}{1 + \alpha}} dp_{k} \wedge dq^{k}, \\
	X_{\alpha_{1}} &:= [X_{\alpha}, Y_{\alpha}] \cr
	& = - \dfrac{1}{2\alpha^{2}} \bigg[ \dfrac{ 1 - \alpha }{(1 + \alpha)} G_{1} |p_{1}|^{-\alpha} |q^{1}|^{\frac{1 - 2\alpha - \alpha^{2}}{1 + \alpha}}\dfrac{\partial}{\partial p_{1}} + |p_{1}|^{1 - \alpha} p_{1}^{-2}|q^{1}|^{\frac{2 - \alpha - \alpha^{2}}{1 + \alpha}} \dfrac{\partial}{\partial q^{1}} \bigg] \cr
	& - \dfrac{1}{2 \alpha^{2}} \sum_{k = 2}^{4}\bigg[ \dfrac{ 1 - \alpha }{(1 + \alpha)} G_{k} p_{k}^{-\alpha} |q^{k}|^{\frac{1 - 2\alpha - \alpha^{2}}{1 + \alpha}}\dfrac{\partial}{\partial p_{k}} + |p_{k}|^{1 - \alpha} p_{k}^{-2}|q^{k}|^{\frac{2 - \alpha - \alpha^{2}}{1 + \alpha}} \dfrac{\partial}{\partial q^{k}} \bigg],
	\end{align*}}
where  $ G_{i} = sgn(p_{i})sgn(q^{i}), \quad i= 1,2,3,4.$ 

We notice that $  X_{\alpha_{1}}$ satisfies the relation   
{\small\begin{equation*} 
	\iota_{_{X_{\alpha_{1}}}} \omega_{\alpha}= -dL_{\alpha}, 
	\end{equation*}}
where $\iota_{_{X_{\alpha_{1}}}} \omega_{\alpha}$ is  the interior product of $ \omega_{\alpha}$ with respect to the  $\al$-vector fied $X_{\alpha_{1}}.$  Since
  $X_{\alpha_{1}}$ is a dynamical symmetry, $i.e.,$ $[X_{\alpha}, X_{\alpha_{1}}] = 0,$  $L_{\alpha}$ is a first integral, also called  a constant of motion. Thus, we arrive at the following property:

\begin{proposition}\label{prop3}
	The $\al$-vector field $X_{\alpha_{1}}$ is an  infinitesimal Noether symmetry.
\end{proposition}
\textbf{Proof. }

We have:
\begin{equation} \label{gs}
\mathcal{L}_{X_{\alpha_{1}}}\omega_{\alpha} = d\iota_{X_{\alpha_{1}}}\omega_{\alpha} +  \iota_{X_{\alpha_{1}}} d\omega_{\alpha} = d\iota_{X_{\alpha_{1}}}\omega_{\alpha} = -d^{2} L_{\alpha} = 0.
\end{equation}
Since $X_{\alpha_{1}}$ is a dynamical symmetry, then 
\begin{equation} \label{hs}
\mathcal{L}_{X_{\alpha_{1}}}H_{\alpha}= X_{\alpha_{1}}(H_{\alpha})  = 0.
\end{equation}
Equations \eqref{gs} and \eqref{hs} show that $X_{\alpha_{1}}$ is both an infinitesimal geometric symmetry, i.e., leaving invariant the geometric structure (the symplectic form $\omega_{\alpha}$),  and an infinitesimal Hamiltonian symmetry leaving invariant the dynamics (the Hamiltonian function $H_{\alpha}$). Hence, $X_{\alpha_{1}}$ is an  infinitesimal Noether symmetry.
$\hfill{\square}$

In the sequel, we consider the  following  $\al$-Poisson bivector
\begin{equation*} \label{poi}
\mathcal{P}_{\alpha_{1}} = p_{1}^{3}|q^{1}|^{\frac{ \alpha - 1}{1 + \alpha}} \dfrac{\partial}{\partial p_{1}} \wedge  \dfrac{\partial}{\partial q^{1}} - \sum_{k = 2}^{4} p_{k}^{3}|q^{k}|^{\frac{ \alpha - 1}{1 + \alpha}} \dfrac{\partial}{\partial p_{k}} \wedge  \dfrac{\partial}{\partial q^{k}}
\end{equation*}
and define the $\al$-deformed  Poisson bracket   
\begin{equation*} \label{pb1}
\{f,g\}_{\alpha_{1}} :=  p_{1}^{3}|q^{1}|^{\frac{ \alpha - 1}{1 + \alpha}}\bigg(\dfrac{\partial{f}}{\partial{p_{1}}}\dfrac{\partial{g}}{\partial{q^{1}}} - \dfrac{\partial{f}}{\partial{q^{1}}}\dfrac{\partial{g}}{\partial{p_{1}}}\bigg) -  \sum_{k=2}^{4}p_{k}^{3}|q^{k}|^{\frac{ \alpha - 1}{1 + \alpha}} \bigg(\dfrac{\partial{f}}{\partial{p_{k}}}\dfrac{\partial{g}}{\partial{q^{k}}} - \dfrac{\partial{f}}{\partial{q^{k}}}\dfrac{\partial{g}}{\partial{p_{k}}}\bigg), 
\end{equation*}
with respect to the $\al$-symplectic form $\omega_{\alpha_{1}}$.

Thus,  the vector field $X_{\alpha}$ is a bi-Hamiltonian vector field  with respect to  $(\omega_{\alpha},\omega_{\alpha_{1}}), $  {\it i.e.,}
{\small\begin{equation*}
	\iota_{_{X_{\alpha}}} \omega_{\alpha} = - dH_{\alpha} \quad \mbox{and} \quad \iota_{X_{\alpha}} \omega_{\alpha_{1}}  = - d\tilde{L}_{\alpha}, \quad X_{\alpha} =  \{ H_{\alpha},.\}_{\alpha} =  \{ \tilde{L}_{\alpha} ,.\}_{\alpha_{1}},
	\end{equation*}}
where 
\begin{equation*}  \tilde{L}_{\alpha} = \displaystyle\sum_{\mu = 1}^{4} |q^{\mu}|^{\frac{1 - \alpha}{1 + \alpha}} p_{\mu}^{-1}    \end{equation*}
are first integrals for $X_{H_{\alpha}}.$

Therefore, the associated $\al$-recursion operator $	T_{\alpha}$ is given by: 
{\small\begin{align*}
	T_{\alpha} & := \mathcal{P}_{\alpha_{1}}\circ \mathcal{P}_{\alpha}^{-1} \cr
	& = \bigg(p_{1}^{3}|q^{1}|^{\frac{ \alpha - 1}{1 + \alpha}} \dfrac{\partial}{\partial p_{1}} \wedge  \dfrac{\partial}{\partial q^{1}} - \sum_{k = 2}^{4} p_{k}^{3}|q^{k}|^{\frac{ \alpha - 1}{1 + \alpha}} \dfrac{\partial}{\partial p_{k}} \wedge  \dfrac{\partial}{\partial q^{k}}  \bigg) \circ \bigg(  \sum_{\mu = 1}^{4} \alpha^{2}|p_{\mu}|^{(\alpha - 1)}|q^{\mu}|^{(\alpha - 1)}dp_{\mu}\wedge dq^{\mu}      \bigg) \cr
	 & = \alpha^{2} p_{1}^{3}|p_{1}|^{(\alpha -1)}  |q^{1}|^{\frac{-2 + \alpha^{2} + \alpha}{ 1 + \alpha}} \dfrac{\partial}{\partial p_{1}} \otimes dp_{1} -  \alpha^{2} \sum_{k = 2}^{4} p_{k}^{3}|p_{k}|^{(\alpha -1)}  |q^{k}|^{\frac{-2 + \alpha^{2} + \alpha}{ 1 + \alpha}}  \dfrac{\partial}{\partial p_{k}} \otimes dp_{k} \cr
	& + \alpha^{2} p_{1}^{3}|p_{1}|^{(\alpha -1)}  |q^{1}|^{\frac{-2 + \alpha^{2} + \alpha}{ 1 + \alpha}} \dfrac{\partial}{\partial q^{1}} \otimes dq^{1} -  \alpha^{2} \sum_{k = 2}^{4} p_{k}^{3}|p_{k}|^{(\alpha -1)}  |q^{k}|^{\frac{-2 + \alpha^{2} + \alpha}{ 1 + \alpha}}   \dfrac{\partial}{\partial q^{k}} \otimes dq^{k},
	\end{align*}}
providing the constants of motion 
{\small\begin{equation*}
	Tr(T_{\alpha}^{h})  = 2^{h}\alpha^{2h}\Bigg\{ \bigg( p_{1}^{3}|p_{1}|^{(\alpha -1)}  |q^{1}|^{\frac{-2 + \alpha^{2} + \alpha}{ 1 + \alpha}}\bigg)^{h} + (-1)^{h}\bigg(  \sum_{k = 2}^{4} p_{k}^{3}|p_{k}|^{(\alpha -1)}  |q^{k}|^{\frac{-2 + \alpha^{2} + \alpha}{ 1 + \alpha}}   \bigg)^{h} \Bigg\}, \ h\in \mathbb{N}.
	\end{equation*}}

This work  can  be considered as an $\al$-deformed case of  previous investigations \cite{tak},  \cite{hkn1}. The only difference  resides in the fact that we use here the method of Noether symmetry to obtain the integrals of motion instead of the method of Hamilton-Jacobi separability developed in \cite{tak} \cite{hkn1}, and \cite{hkn2}. 

\section{$\alpha$-deformed  Einstein field equation}\label{sec4}
In this section, we investigate the $\al$-deformed 
 solutions of the Einstein field equation in the Schwarzschild and Friedmann-Lemaître-Robertson-Walker (FLRW) metrics.
We consider the $\al$-deformed  Einstein field equation shortly written in the tensor form as:
\[
G_{\al} + \Lambda g_{\al} = \kappa T_{\al},
\]
where the tensor 
\begin{equation}\label{Eins} 
G_{\al} =  R_{\al} - \dfrac{1}{2}g_{\al}\mathbf{R}_{\al}
\end{equation} 
is the $\al$-deformed   Einstein tensor, the constant $\Lambda$ is  the cosmological constant, $\kappa$ is a constant;  $T_{\al}$  and $R_{\al}$ are the  $\al-$deformed tress-energy tensor and Ricci tensor measuring the geodesic deviation, respectively. $g_{\al}$ is the $\al-$deformed metric tensor telling us how to measure distances and time, and $\mathbf{R}_{\al},$ analogous to Gaussian curvature $\kappa$ for surfaces, is the $\al-$deformed scalar curvature measuring the spacetime curvature.
The right hand side of the Einstein field equation is the matter content of spacetime, in here we can put electromagnetic fields, fluids, scalar fields, etc.
  Therefore, Einstein equations express the compromise between spacetime geometry and the existing matter, in such a way that each one influences the other, being the curvature the manifestation of the massive content of that spacetime. The main difference with respect to the old approach is that matter evolves no longer through a static spatial scenario, where all clocks in universe agree in their time measurements, but rather now spacetime is an active actor that affects matter dynamics; and in turn the matter content, through the energy-momentum tensor $T_{\al}$, determines how the geometry is. Moreover, Einstein equations are not linear, where by linear effects we understand those that are proportional to the causes, those where small variations in initial conditions lead to small changes in the response, not much different than the former. In nonlinear processes, this is not the case.

\subsection{ Recursion operator of the $\al$-deformed Schwarzschild metric }

The Schwarzschild metric is the simplest one among the  particular solutions of the Einstein field equation.\\
Here, we consider the following $\alpha$-Schwarzschild metric
\begin{align*}
d_{\alpha}s^{2} & = - \bigg(1 - \dfrac{2M}{q^{2}}\bigg)(q^{1})^{2(\alpha -1)} (dq^{1})^{2} + \bigg(1 - \dfrac{2M}{q^{2}}\bigg)^{-1}(q^{2})^{2(\alpha -1)}(dq^{2})^{2} \cr
& + (q^{2})^{2}(q^{3})^{2(\alpha -1)}(dq^{3})^{2} + (q^{2})^{2}(q^{4})^{2(\alpha -1)}\sin^{2}q^{3}(dq^{4})^{2},\end{align*}
where $
t = q^{1},\ r = q^{2},\ \theta = q^{3} ,\ \phi = q^{4} ,$
$M$ is a positive constant representing the mass of the black hole,
$ t \in ( -\infty , \infty),$  $r \in (2M , \infty),$  $\theta \in (0 , \pi),$ and $\phi \in (0, 2\pi)$. 

The metric is defined on a manifold
\begin{equation} \label{mani}
\mathcal{Q} = \{ (q^{1},q^{2},q^{3},q^{4})|\ 0\neq q^{1} \in ( -\infty , \infty) , q^{2} \in (2M , \infty) , 0\neq q^{3} \in (0 , \pi), \ \mbox{and} \ 0\neq q^{4} \in (0, 2\pi)\}.
\end{equation}
For ${\alpha}=1,$ we recover the metric  found by Karl Schwarzschild in 1916 as  the simplest possible solution of Einstein equations, which was of huge usefulness to derive some effects or predictions as well as to lead to the prediction of the existence of black holes, very compact massive objects.
It was the only spherically symmetric, vacuum solution of Einstein equations. It describes the geometry exterior to any spherical collapsing body. And even in the case of nonspherical collapse, perturbative treatment of spherical collapse indicates that a blackhole will form and the spacetime singularity resulting from the gravitational collapse will be hidden within the black hole \cite{breton}.

For our purpose, let us consider the phase space $\mathcal{T}^{\ast} \mathcal{Q}  \ni (q,p), q \in \mathcal{Q} ,$ and the $\al$-Hamiltonian function

\begin{align*}
H_{S\alpha} & = -\dfrac{1}{2} \bigg(1 - \dfrac{2M}{q^{2}}\bigg)^{-1}(q^{1})^{2(1 -\alpha)}p_{1}^{2} + \dfrac{1}{2}\bigg(1 - \dfrac{2M}{q^{2}}\bigg)(q^{2})^{2( 1 -\alpha)}p_{2}^{2} \cr
&  + \dfrac{1}{2(q^{2})^{2}} (q^{3})^{2( 1 -\alpha)}p_{3}^{2} +  \dfrac{1}{2(q^{2})^{2}\sin^{2}q^{3}} (q^{4})^{2( 1 -\alpha)}p_{4}^{2}.
\end{align*}
The $\al$-Hamiltonian vector field of $	H_{S\alpha}$ in $\al$-Schwarzschild metric with respect to the canonical 	$\al$-symplectic structure $\omega_{\alpha} = \displaystyle\sum_{\mu = 1}^{4} \al^{2} |p_{\mu}|^{(\alpha - 1)}|q^{\mu}|^{(\alpha - 1)}dp_{\mu}\wedge dq^{\mu} $ is given by
{\small\begin{align*}
	X_{S\alpha} &:= \{ H_{S\alpha}, .\}_{\alpha} = \sum_{\mu = 1}^{4}\al^{-2}|p_{\mu}|^{( 1- \alpha)}|q^{\mu}|^{(1 - \alpha )} \Bigg(\dfrac{\partial{H_{\alpha}}}{\partial{p_{\mu}}}\dfrac{\partial}{\partial{q^{\mu}}} - \dfrac{\partial{H_{\alpha}}}{\partial{q^{\mu}}}\dfrac{\partial}{\partial{p_{\mu}}}\Bigg) \cr
	& = \sum_{\mu = 1}^{4}\al^{-2}|p_{\mu}|^{( 1- \alpha)}|q^{\mu}|^{(1 - \alpha )} \bigg( V_{\mu}\dfrac{\partial}{\partial{q^{\mu}}} + U_{\mu}\dfrac{\partial}{\partial{p_{\mu}}}  \bigg),
	\end{align*}}
where  
{\small\begin{align*}
	V_{1} & = -\bigg(1 - \dfrac{2M}{q^{2}}\bigg)^{-1} \eta_{1}, \quad V_{2} = \bigg(1 - \dfrac{2M}{q^{2}}\bigg) \eta_{2}, \
	V_{3}  = \dfrac{1}{ (q^{2})^{2}} \eta_{3}, \ V_{4} = \dfrac{1}{ (q^{2})^{2} \sin^{2}q^{3}} \eta_{4}, \\
	U_{1} & = (1- \alpha ) \bigg(1 - \dfrac{2M}{q^{2}}\bigg)^{-1} \zeta_{1}, \ U_{3} =  - \bigg( \dfrac{ 1 - \alpha}{(q^{2})^{2}}  \zeta_{3} - \dfrac{\cos q^{3}}{(q^{2})^{2} \sin^{3}q^{3}}  \eta_{4} p_{4} \bigg), \   U_{4} = - \dfrac{ 1 - \alpha}{(q^{2})^{2}\sin^{2}q^{3}}  \zeta_{4},  \\
	U_{2} & = - \bigg\{\dfrac{M}{ (q^{2})^{2}}\bigg(1 - \dfrac{2M}{q^{2}}\bigg)^{-2} \eta_{1}p_{1} + (1 -\alpha)\bigg(1 - \dfrac{2M}{q^{2}}\bigg) \zeta_{2} 
	+ \dfrac{M}{ (q^{2})^{2}}\eta_{2}p_{2} - \dfrac{1}{ (q^{2})^{3}} \eta_{3}p_{3} \cr 
	& -  \dfrac{1}{ (q^{2})^{3}\sin^{2}q^{3}} \eta_{4}p_{4} \bigg\},
	\end{align*}}
with $\eta_{\nu} = (q^{\nu})^{2(1-\alpha)}p_{\nu},$ and $ \zeta_{\nu} = (q^{\nu})^{(1-2\alpha)}p_{\nu}^{2}, \ \nu = 1,2,3,4. $

Then, we get in $\al$-Schwarzschild metric, the $\al$-Christoffel symbols $(\Gamma^{k}_{ij})_{\alpha},$ the components of the $\al$-Riemann $(R_{ijkl})_{\alpha}$ and $\al$-Ricci tensors $(R_{ii})_{\alpha},$ the $\al$-Ricci scalar $\mathbf{R},$  and the components of the $\al$-Einstein tensor $(G_{ij})_{\alpha}$,\ $i,j,k,l = 1,2,3,4$, see  Appendix.

Note that the components of  defined geometric objects  are obtained in the usual undeformed Schwarzschild metric by setting $\al = 1.$

Now, we consider the Hamilton-Jacobi equation for the $\al$-Hamiltonian function $H_{S\alpha }$
\begin{align*} 
E_{S} & = H_{S\alpha} \bigg( q, \dfrac{\partial W}{\partial q}\bigg) \cr
& = -\dfrac{1}{2} \bigg(1 - \dfrac{2M}{q^{2}}\bigg)^{-1}(q^{1})^{2(1 -\alpha)} \bigg(\dfrac{\partial W}{\partial q^{1}}\bigg)^{2} + \dfrac{1}{2}\bigg(1 - \dfrac{2M}{q^{2}}\bigg)(q^{2})^{2( 1 -\alpha)}\bigg(\dfrac{\partial W}{\partial q^{2}}\bigg)^{2} \cr
&  + \dfrac{1}{2(q^{2})^{2}} (q^{3})^{2( 1 -\alpha)}\bigg(\dfrac{\partial W}{\partial q^{3}}\bigg)^{2} +  \dfrac{1}{2(q^{2})^{2}\sin^{2}q^{3}} (q^{4})^{2( 1 -\alpha)}\bigg(\dfrac{\partial W}{\partial q^{4}}\bigg)^{2},
\end{align*}
where $E_{S}$ is a constant and $W = \displaystyle\sum_{\mu = 1}^{4}W_{\mu}(q_{\mu})$ is the generating function. In particular, we put $ W_{1} = \dfrac{a}{\alpha} |q^{1}|^{\alpha}$, where $a$ is a constant. This equation is a type of separation of variables; then, the above Hamilton-Jacobi equation becomes 
{\small \begin{align} \label{Sep1}
	& 2  E_{S} (q^{2})^{2}  + \bigg(1 - \dfrac{2M}{q^{2}}\bigg)^{-1} (q^{2})^{2} a^{2} -  \bigg(1 - \dfrac{2M}{q^{2}}\bigg)(q^{2})^{2( 2 -\alpha)}\bigg(\dfrac{dW_{2}}{dq^{2}}\bigg)^{2} \cr
	& =    (q^{3})^{2( 1 -\alpha)}\bigg(\dfrac{d W_{3}}{d q^{3}}\bigg)^{2} +  \dfrac{1}{\sin^{2}q^{3}} (q^{4})^{2( 1 -\alpha)}\bigg(\dfrac{dW_{4}}{ dq^{4}}\bigg)^{2},
	\end{align}}
 which can be rewritten through a constant $K$ as: 
{\small \begin{align}
	K & = 2  E_{S} (q^{2})^{2}  + \bigg(1 - \dfrac{2M}{q^{2}}\bigg)^{-1} (q^{2})^{2} a^{2} -  \bigg(1 - \dfrac{2M}{q^{2}}\bigg)(q^{2})^{2( 2 -\alpha)}\bigg(\dfrac{d W_{2}}{d q^{2}}\bigg)^{2} \label{sp4}\\
	K & = (q^{3})^{2( 1 -\alpha)}\bigg(\dfrac{d W_{3}}{dq^{3}}\bigg)^{2} +  \dfrac{1}{\sin^{2}q^{3}} (q^{4})^{2( 1 -\alpha)}\bigg(\dfrac{d W_{4}}{d q^{4}}\bigg)^{2}.
	\end{align}}
From the above, we set: 
{\small \begin{align}
	\bigg( K - (q^{3})^{2( 1 -\alpha)}\bigg(\dfrac{d W_{3}}{d q^{3}}\bigg)^{2} \bigg) \sin^{2}q^{3} & = G \label{sp5} \\
	(q^{4})^{2( 1 -\alpha)}\bigg(\dfrac{d W_{4}}{d q^{4}}\bigg)^{2} & = G. \label{sp3}
	\end{align}}
and obtain
{\small \begin{equation*}
	W_{4} = \dfrac{\sqrt{G}}{\alpha} | q^{4}|^{\alpha - 1}q^{4} + A,
	\end{equation*}}
where $A$ is a constant. 

We put the solutions of equations \eqref{sp4} and \eqref{sp5} in the form:  
{\small \begin{equation*}
	W_{2} =  W_{2}(q^{2}, E_{S}, a, K), \quad W_{3} =  W_{3}(q^{3}, K, G).
	\end{equation*}}
Then, a generating function $W$ takes the form:
{\small \begin{equation*}
	W = \dfrac{a}{\alpha} |q^{1}|^{\alpha}   +  W_{2}(q^{2}, E_{S}, a, K) +   W_{3}(q^{3}, K, G) +  \dfrac{\sqrt{G}}{\alpha} | q^{4}|^{\alpha - 1}q^{4}  + A. 
	\end{equation*}}
Now, we consider the canonical system $(Q, P)$, where 
{\small \begin{align*}
	& Q^{1} = E_{S}, \ Q^{2} = a, \  Q^{3} = K, \ Q^{4} = \sqrt{G}, \\
	& P_{1} := -\dfrac{\partial W}{\partial Q^{1}} = -\dfrac{\partial W_{2}}{\partial Q^{1}}, \quad  P_{2} := -\dfrac{\partial W}{\partial Q^{2}} = -\dfrac{a}{\alpha}(q^{1})^{\alpha} - \dfrac{\partial W_{2}}{\partial Q^{2}},\cr
	&  P_{3} := -\dfrac{\partial W}{\partial Q^{3}} = -\dfrac{\partial W_{2}}{\partial Q^{3}} - \dfrac{\partial W_{3}}{\partial Q^{3}}  , \ \mbox{and} \
	 P_{4} := -\dfrac{\partial W}{\partial Q^{4}} = -\dfrac{\partial W_{4}}{\partial Q^{4}} - \dfrac{\partial W_{3}}{\partial Q^{4}} = - \dfrac{1}{\alpha}|q^{4}|^{\alpha -1}q^{4} - \dfrac{\partial W_{3}}{\partial Q^{4}}.
	\end{align*}}

In this new canonical system, we define the following Poisson bracket
{\small\begin{equation} \label{conPoi2}
	\{f,g\}_{\alpha}= \sum_{\mu = 1}^{4}\al^{-2}|P_{\mu}|^{( 1- \alpha)}|Q^{\mu}|^{(1 - \alpha )} \Bigg(\dfrac{\partial{f}}{\partial{P_{\mu}}}\dfrac{\partial{g}}{\partial{Q^{\mu}}} - \dfrac{\partial{f}}{\partial{Q^{\mu}}}\dfrac{\partial{g}}{\partial{P_{\mu}}}\Bigg), 
	\end{equation}}
with respect to the  $\al$-symplectic form 
{\small\begin{equation*}
	\omega_{\alpha} = \sum_{\mu = 1}^{4} \al^{2} |P_{\mu}|^{(\alpha - 1)}|Q^{\mu}|^{(\alpha - 1)}dP_{\mu}\wedge dQ^{\mu}.
	\end{equation*}}

Then,
the $\al$-Hamiltonian vector field takes the form:
{\small\begin{equation*}
	X_{S\alpha} :=  \{H_{S\al},.\}_{\alpha}= - \al^{-2}|P_{1}|^{( 1- \alpha)}|Q^{1}|^{(1 - \alpha )} \dfrac{\partial{}}{\partial{P_{1}}}.
	\end{equation*}}
Now, we consider a $(1,1)$-tensor field $T_{S\alpha}$ as 
\begin{equation*}
T_{S\alpha} = \sum_{\mu = 1}^{4} |Q^{\mu}|^{\alpha - 1} Q^{\mu}\bigg( \dfrac{\partial{}}{\partial{P_{\mu}}}\otimes dP_{\mu} + \dfrac{\partial{}}{\partial{Q^{\mu}}}\otimes dQ^{\mu}\bigg).
\end{equation*}
We can put $Q^{\mu} = x_{\mu}$ and $P_{\mu} = x_{\mu + n},$ where $n =4$ in this case and $\mu = 1,2,3,4$.
Then, by Lemma \ref{lem1}, $T_{S\alpha}$ satisfies $\mathcal{L}_{X_{S\alpha}}T_{S\alpha} = 0$, $\mathcal{N}_{T_{S\alpha}} = 0,$ and $deg Q^{\mu} = 2$. Hence,  $T_{S\alpha}$ is an $\al$-recursion operator of $X_{S\alpha}$. The constants of motion $Tr(	T_{\alpha}^{l}) \ (l \in \mathbb{N})$ of the $\al$-Hamiltonian vector field $X_{S\alpha}$ for the $\alpha$-Schwarzschild metric are finally obtained as:
\[
Tr(	T_{F\alpha}^{l}) = 2((Q^{1})^{l} + (Q^{2})^{l} + (Q^{3})^{l} + (Q^{4})^{l}), \quad l \in \mathbb{N}.
\]

\subsection{Recursion operator in  $\alpha$-FLRW metric}
Now, we consider the following $\alpha$-Friedmann-Lema\^itre-Robertson-Walker (FLRW) metric:
\begin{align*}
d_{\alpha}s^{2} & = -|q^{1}|^{2(\alpha - 1)}(dq^{1})^{2} + R^{2}(q^{1})\bigg\{\dfrac{|q^{2}|^{2(\alpha - 1)}}{1 - k(q^{2})^{2}}(dq^{2})^{2}  + (q^{2})^{2}\bigg(|q^{3}|^{2(\alpha - 1)}(dq^{3})^{2} \cr
& + |q^{4}|^{2(\alpha - 1)}\sin^{2}q^{3}(dq^{4})^{2}\bigg) \bigg\}
\end{align*}
defined on the same manifold $\mathcal{Q}$ \eqref{mani}, where 
$R(q^{1})$ is a scale factor and $k$ is a constant representing the curvature of the space. Considering the $\al$-Hamitonian function 
{\small\begin{align*}
	H_{F\alpha} = -\dfrac{1}{2} (q^{1})^{2(1 -\alpha)}p_{1}^{2} +\dfrac{1 - k(q^{2})^{2}}{2R^{2}(q^{1})}(q^{2})^{2(1 -\alpha)}p_{2}^{2} + \dfrac{(q^{3})^{2(1 -\alpha)}}{2(q^{2})^{2}R^{2}(q^{1})}p_{3}^{2} +  \dfrac{(q^{4})^{2(1 -\alpha)}}{2(q^{2})^{2}R^{2}(q^{1}) \sin^{2}(q^{3})}p_{4}^{2},
	\end{align*}}
we obtain the following $\al$-Hamiltonian vector field 
{\small\begin{align*}
	X_{F\alpha} = \sum_{\mu = 1}^{4}\al^{-2}|p_{\mu}|^{( 1- \alpha)}|q^{\mu}|^{(1 - \alpha )} \bigg( \tilde{V}_{\mu}\dfrac{\partial}{\partial{q^{\mu}}} + \tilde{U}_{\mu}\dfrac{\partial}{\partial{p_{\mu}}}  \bigg),
	\end{align*}}  
with respect to the $\al$-symplectic structure $\omega_{\alpha} = \displaystyle\sum_{\mu = 1}^{4}\al^{2} |p_{\mu}|^{(\alpha - 1)}|q^{\mu}|^{(\alpha - 1)}dp_{\mu}\wedge dq^{\mu},$
where 
{\small\begin{align*} 
	& \tilde{V}_{1} = \eta_{1}, \quad \tilde{V}_{2} = \dfrac{1 - k(q^{2})^{2}}{2R^{2}(q^{1})}\eta_{2}, \quad  \tilde{V}_{3} =  \dfrac{1}{(q^{2})^{2}R^{2}(q^{1})}\eta_{3}, \quad \tilde{V}_{4} = \dfrac{1}{(q^{2})^{2}R^{2}(q^{1}) \sin^{2}(q^{3})}\eta_{4},\\
	& \tilde{U}_{1} = (1-\alpha)\zeta_{1} + \dfrac{1}{R^{3}(q^{1})}\bigg((1 - k(q^{2})^{2})\eta_{2}p_{2} + \dfrac{1}{(q^{2})^{2}}\eta_{3}p_{3} +\dfrac{1}{\sin^{2}q^{3}}\eta_{4}p_{4}  \bigg)\dfrac{dR(q^{1})}{dq^{1}}, \\
	& \tilde{U}_{2} = - \dfrac{p^{2}_{2}}{R^{2}(q^{1})} \bigg(-kq^{2}\eta_{2}p_{2} + (1 - \alpha)(1 - k(q^{2})^{2})\zeta_{2}\bigg) + \dfrac{1}{(q^{2})^{3}R^{2}(q^{1})}\eta_{3}p_{3} + \dfrac{1}{(q^{2})^{3}R^{2}(q^{1})\sin^{2}q^{3}}\eta_{4}p_{4}, \\
	&  \tilde{U}_{3} = - \dfrac{(1 - \alpha)}{(q^{2})^{2}R^{2}(q^{1})}\zeta_{3} + \dfrac{ \cos q^{3}}{(q^{2})^{2}R^{2}(q^{1})sin^{3}q^{3}}\eta_{4}p_{4}, \quad \tilde{U}_{4} =  -\dfrac{(1 - \alpha)}{(q^{2})^{2}R^{2}(q^{1})sin^{2}q^{3}}\zeta_{4},
	\end{align*}}  
with $\eta_{\nu} = (q^{\nu})^{2(1-\alpha)}p_{\nu},$ $ \zeta_{\nu} = (q^{\nu})^{(1-2\alpha)}p_{\nu}^{2},$ and $\nu = 1,2,3,4. $

Here, we perform the computation of the $\al$-Christoffel symbols, the components of the $\al$-Riemann and $\al$-Ricci tensors, the $\al$-Ricci scalar, and the components of the Einstein tensor, 
 see Appendix.

Remark that for $\al = 1,$ we recover the components of these geometric objects in the usual FLRW metric as expected.   

The Hamiltonian-Jacobi equation takes here the form:
{\small\begin{align*} 
	2E_{F} & =  - (q^{1})^{2(1 -\alpha)} \bigg(\dfrac{\partial W}{\partial q^{1}}\bigg)^{2} +\dfrac{1 - k(q^{2})^{2}}{R^{2}(q^{1})}(q^{2})^{2(1 -\alpha)}\bigg(\dfrac{\partial W}{\partial q^{2}}\bigg)^{2} \cr
	& + \dfrac{(q^{3})^{2(1 -\alpha)}}{(q^{2})^{2}R^{2}(q^{1})}\bigg(\dfrac{\partial W}{\partial q^{3}}\bigg)^{2} +  \dfrac{(q^{4})^{2(1 -\alpha)}}{(q^{2})^{2}R^{2}(q^{1}) \sin^{2}(q^{3})}\bigg(\dfrac{\partial W}{\partial q^{4}}\bigg)^{2},
	\end{align*}} 
where $E_{F}$ is a constant and $W = \displaystyle\sum_{\mu = 1}^{4}W_{\mu}(q_{\mu})$ is the generating function.
The above equation can be rewritten as 
{\small\begin{align*} 
	2E_{F}R^{2}(q^{1}) + (q^{1})^{2(1 -\alpha)}R^{2}(q^{1})\bigg(\dfrac{dW_{1}}{ dq^{1}}\bigg)^{2} & =  (1 - k(q^{2})^{2})(q^{2})^{2(1 -\alpha)}\bigg(\dfrac{d W_{2}}{d q^{2}}\bigg)^{2} + \dfrac{(q^{3})^{2(1 -\alpha)}}{(q^{2})^{2}}\bigg(\dfrac{d W_{3}}{d q^{3}}\bigg)^{2} \cr
	& +  \dfrac{(q^{4})^{2(1 -\alpha)}}{(q^{2})^{2} \sin^{2}(q^{3})}\bigg(\dfrac{dW_{4}}{d q^{4}}\bigg)^{2},
	\end{align*}} 
which is of a type of separation of variables. Thus, 
we can also express them via a constant $K$ as: 
{\small\begin{align} 
	& K = 2E_{F}R^{2}(q^{1}) + (q^{1})^{2(1 -\alpha)}R^{2}(q^{1})\bigg(\dfrac{dW_{1}}{ dq^{1}}\bigg)^{2}, \label{W1}\\
	& K =  (1 - k(q^{2})^{2})(q^{2})^{2(1 -\alpha)}\bigg(\dfrac{d W_{2}}{d q^{2}}\bigg)^{2} + \dfrac{(q^{3})^{2(1 -\alpha)}}{(q^{2})^{2}}\bigg(\dfrac{d W_{3}}{d q^{3}}\bigg)^{2} +  \dfrac{(q^{4})^{2(1 -\alpha)}}{(q^{2})^{2} \sin^{2}(q^{3})}\bigg(\dfrac{dW_{4}}{d q^{4}}\bigg)^{2}.\label{Ja1}
	\end{align}} 
Moreover, from equation \eqref{Ja1}, we get
{\small\begin{align} \label{Ja2}
	(1 - k(q^{2})^{2})(q^{2})^{2(2 -\alpha)}\bigg(\dfrac{d W_{2}}{d q^{2}}\bigg)^{2} -(q^{2})^{2}K= - (q^{3})^{2(1 -\alpha)}\bigg(\dfrac{d W_{3}}{d q^{3}}\bigg)^{2} -  \dfrac{(q^{4})^{2(1 -\alpha)}}{ \sin^{2}(q^{3})}\bigg(\dfrac{dW_{4}}{d q^{4}}\bigg)^{2}.
	\end{align}} 
Since equation \eqref{Ja2} is of a type of separation of variables, we can introduce a constant $L$ such that 

{\small\begin{align} 
	L & = (q^{2})^{2}K - (1 - k(q^{2})^{2})(q^{2})^{2(2 -\alpha)}\bigg(\dfrac{d W_{2}}{d q^{2}}\bigg)^{2},  \label{W2}\\
	L &= (q^{3})^{2(1 -\alpha)}\bigg(\dfrac{d W_{3}}{d q^{3}}\bigg)^{2} + \dfrac{(q^{4})^{2(1 -\alpha)}}{ \sin^{2}(q^{3})}\bigg(\dfrac{dW_{4}}{d q^{4}}\bigg)^{2} \label{Ja3},
	\end{align}}
and the equation \eqref{Ja3} can be expressed as   
{\small\begin{align*}
	L \sin^{2}(q^{3}) - (q^{3})^{2(1 -\alpha)}\sin^{2}(q^{3})\bigg(\dfrac{d W_{3}}{d q^{3}}\bigg)^{2} = (q^{4})^{2(1 -\alpha)}\bigg(\dfrac{dW_{4}}{d q^{4}}\bigg)^{2}.
	\end{align*}}
Setting
{\small\begin{align} 
	G & = L \sin^{2}(q^{3}) - (q^{3})^{2(1 -\alpha)}\sin^{2}(q^{3})\bigg(\dfrac{d W_{3}}{d q^{3}}\bigg)^{2},  \label{W3} \\
	G & =  (q^{4})^{2(1 -\alpha)}\bigg(\dfrac{dW_{4}}{d q^{4}}\bigg)^{2}, \label{Ja4}
	\end{align}}
 we can formulate the solutions of the equations \eqref{W1}, \eqref{W2}, and \eqref{W3} as: 
{\small\begin{align*} 
	W_{1} = W_{1}(q^{1};E_{F},K), \ W_{2} = W_{2}(q^{2};K,L), \ W_{3} = W_{3}(q^{3};L,G).
	\end{align*}}
From \eqref{Ja4}, we obtain  
{\small\begin{align*} 
	W_{4} = \dfrac{\sqrt{G}}{\alpha} | q^{4}|^{\alpha - 1}q^{4} + C,
	\end{align*}}
where $C$ is a constant, 
and, hence,
{\small\begin{align*} 
	W = W_{1}(q^{1};E_{F},K) + W_{2}(q^{2};K,L) + W_{3}(q^{3};L,G) + \dfrac{\sqrt{G}}{\alpha} | q^{4}|^{\alpha - 1}q^{4} + C.
	\end{align*}}

Considering now the canonical system $(Q, P)$, where 
{\small \begin{align*}
	& Q^{1} = E_{F}, \ Q^{2} = K, \  Q^{3} = \sqrt{L}, \ Q^{4} = \sqrt{G}, \\
	& P_{1} := -\dfrac{\partial W}{\partial Q^{1}} = -\dfrac{\partial W_{1}}{\partial Q^{1}}, \quad  P_{2} := -\dfrac{\partial W}{\partial Q^{2}} = -\dfrac{\partial W_{1}}{\partial Q^{2}} - \dfrac{\partial W_{2}}{\partial Q^{2}},\\
	&  P_{3} := -\dfrac{\partial W}{\partial Q^{3}} = -\dfrac{\partial W_{2}}{\partial Q^{3}} - \dfrac{\partial W_{3}}{\partial Q^{3}}, \ \mbox{and} \  P_{4} := -\dfrac{\partial W}{\partial Q^{4}} = -\dfrac{\partial W_{3}}{\partial Q^{4}} - \dfrac{\partial W_{4}}{\partial Q^{4}} = - \dfrac{1}{\alpha}|q^{4}|^{\alpha -1}q^{4} - \dfrac{\partial W_{3}}{\partial Q^{4}},
	\end{align*}}
the $\al$-Hamiltonian vector field $X_{F\alpha}$ and the $(1,1)$-tensor field $	T_{F\alpha}$ are given by
{\small\begin{equation*}
	X_{F\alpha} :=  \{H_{F\al},.\}_{\alpha}= - \al^{-2}|P_{1}|^{( 1- \alpha)}|Q^{1}|^{(1 - \alpha )} \dfrac{\partial{}}{\partial{P_{1}}}, \
	T_{F\alpha} = \sum_{\mu = 1}^{4} |Q^{\mu}|^{\alpha - 1} Q^{\mu}\bigg( \dfrac{\partial{}}{\partial{P_{\mu}}}\otimes dP_{\mu} + \dfrac{\partial{}}{\partial{Q^{\mu}}}\otimes dQ^{\mu} \bigg),
	\end{equation*}}
respectively.

Similarly, by Lemma \ref{lem1}, $T_{F\alpha}$ satisfies $\mathcal{L}_{X_{F\alpha}}T_{F\alpha} = 0$, $\mathcal{N}_{T_{F\alpha}} = 0,$ and $deg Q^{\mu}= 2$. Thus,  $ T_{F\alpha}$ is an  $\al$-recursion operator of $X_{F\alpha},$ and the constants of motion $Tr(	T_{F\alpha}^{l}) \ (l \in \mathbb{N})$ of the $\al$-vector field $X_{F\alpha}$ for the  $\alpha$-FLRW  metric are provided in the form
\[
Tr(	T_{F\alpha}^{l}) = 2((Q^{1})^{l} + (Q^{2})^{l} + (Q^{3})^{l} + (Q^{4})^{l}), \quad l \in \mathbb{N}.
\]

\section{Family of conserved quantities}\label{sec5}
In this section, we consider  the Hamiltonian system $(\mathcal{T}^{\ast}\mathcal{Q}, \omega, Q^{1}),$ for which the $\al$-Hamiltonian function $ H_{\al}$,
the $\al$-vector field $X_{\al}$, the $\al$-symplectic form $ \omega_{\al}$,  the $\al$-bivector field $\mathcal{P_{\al}}$, and the $\al$-recursion operator $T_{\alpha}$ are given in both the  $\al$-Schwarzschild and $\al$-FLRW metrics by: 
$	H_{\al} = Q^{1}> 0,$
{\small\[
	\ X_{\al} = \{H_{\al},.\}_{\alpha}= - \al^{-2}|P_{1}|^{( 1- \alpha)}|Q^{1}|^{(1 - \alpha )} \dfrac{\partial{}}{\partial{P_{1}}},
	\   \omega_{\alpha} = \sum_{\mu = 1}^{4} \al^{2} |P_{\mu}|^{(\alpha - 1)}|Q^{\mu}|^{(\alpha - 1)}dP_{\mu}\wedge dQ^{\mu}, \]
	\[ \mathcal{P_{\al}} = \sum_{\mu = 1}^{4} \al^{-2} |P_{\mu}|^{(1 -\alpha )}|Q^{\mu}|^{(1 - \alpha)} \dfrac{\partial{}}{\partial{P_{\mu}}}\wedge \dfrac{\partial{}}{\partial{Q^{\mu}}}, \ \mbox{and} \ T_{\alpha} = \sum_{\mu = 1}^{4} |Q^{\mu}|^{\alpha - 1} Q^{\mu}\bigg( \dfrac{\partial{}}{\partial{P_{\mu}}}\otimes dP_{\mu} + \dfrac{\partial{}}{\partial{Q^{\mu}}}\otimes dQ^{\mu} \bigg).
	\]}
In the sequel, we introduce  the functions
\begin{align*}
\tilde{	H}_{\al_{j}} = -\sum_{\mu = 1}^{4}\al |Q^{\mu}|^{\al(1 -j ) - 1}Q^{\mu}|P_{\mu}|^{\al -1}P_{\mu}
\end{align*}
and obtain the  vector fields $ 	Z_{\al_{j}} \in \mathcal{T}^{\ast}\mathcal{Q},$

{\small\begin{align*} 
	Z_{\al_{j}} &:= \{\tilde{H}_{\al_{j}}, .\}_{\al} 
	= \sum_{\mu = 1}^{4} |Q^{\mu}|^{-\al j}\bigg((1-j) P_{\mu} \dfrac{\partial{}}{\partial{P_{\mu}}} - Q^{\mu}\dfrac{\partial{}}{\partial{Q^{\mu}}}\bigg).
	\end{align*}}

satisfying the relation
{\small\begin{equation*}
	\iota_{Z_{\al_{j}}}\omega_{\al} = -d\tilde{	H}_{\al_{j}}.
	\end{equation*}}
Then, it is straightforward  to notice that the $\al$-symplectic structure $\omega_{\al}$ generates a set of Hamiltonian systems on the same manifold $\mathcal{T}^{\ast}\mathcal{Q}.$
The Lie bracket  between the vector fields $X_{_{\al_{i}}}$ and $Z_{\al_{j}}$ obeys the relations
{\small \begin{equation} \label{symgr}
	[X_{_{\al_{i}}}, Z_{\al_{j}}] = X_{_{\al_{i + j}}} ,\ [X_{_{\al_{i}}}, X_{_{\al_{i + j}}}] = 0, \ i,j \in \mathbb{N}, \  X_{\al_{0}} = X_{\al},
	\end{equation}
	with 	
	\begin{equation*}	 
	X_{\al_{i +j}} = - \al^{-2}(1 - \al i )[1 - (i +j)\al ] |Q^{1}|^{1 - \al(i +j + 1)  }|P_{1}|^{(1 -\al )}\dfrac{\partial{}}{\partial{P_{1}}} .
	\end{equation*}}
These relations are well diagrammatically represented in   Fig $1$.
In  terms of differential geometry, $Z_{\al_{j}}$  and 	$\tilde{H}_{\al_{j}}$ are called  $\al$-{\it master symmetries} for 	$X_{\al_{i}}$ and   $\al$-{\it master integrals}, respectively. For more details on these symmetries, see \cite{cas,dam,fer2,ra,ra3}.
\begin{figure}[!th]
	\centering
	\includegraphics[width=12cm, height=4cm]{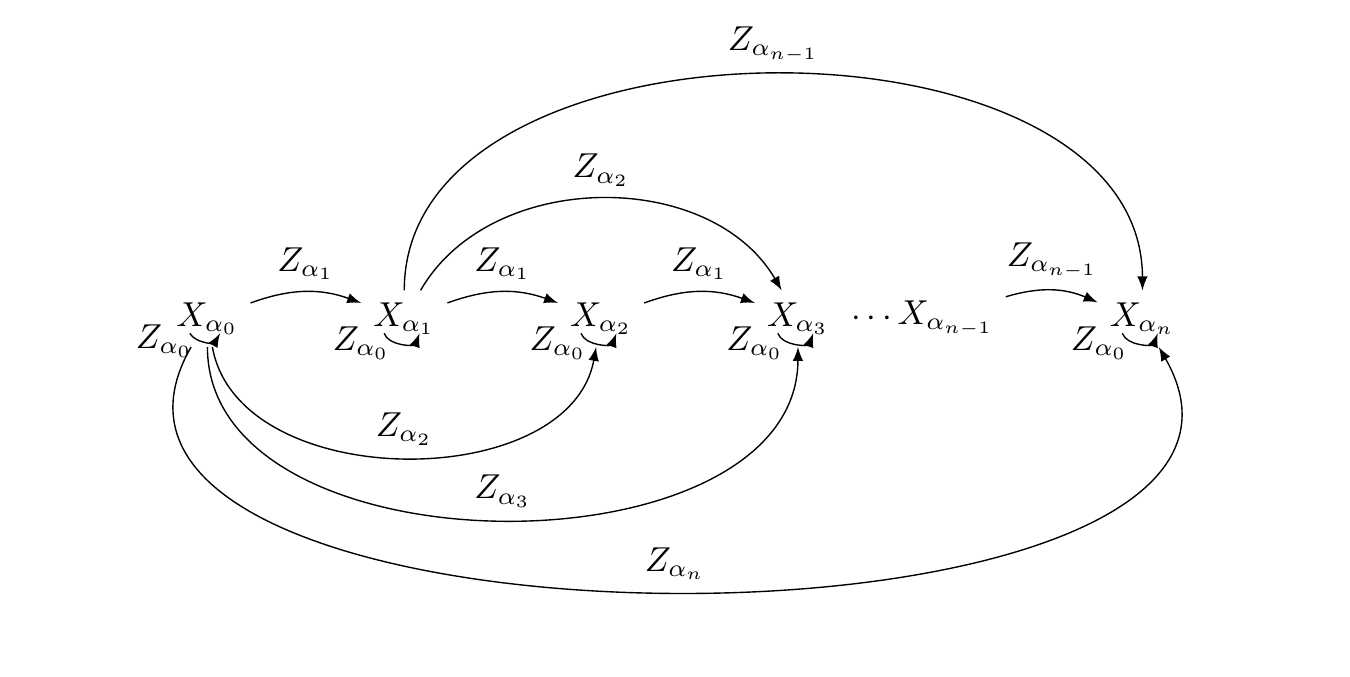}
	\caption{Diagrammatical illustration of equation \eqref{symgr}.}
	\label{Graphs3}
\end{figure}

Thus, we can generate a family  of $\al$-Hamiltonian functions:
\begin{equation*} \label{symv2}
H_{\al_{i + j}} := \{ H_{\al_{i}}, \tilde{	H}_{\al_{j}}\} = (1 - \al i ) (Q^{1})^{1 -\al (i +j) } , \ \mbox{with} \   H_{\al_{0}}=  H_{\al}, \ i, j \in \mathbb{N}.	
\end{equation*}
The $\al$-recursion operator $T_{\al}$
can be written as:
\[
T_{\al} = \mathcal{P}_{\al_{1}}\circ \mathcal{P}_{\al}^{-1},
\]
where 
\begin{equation*}
\mathcal{	P}_{\alpha_{1}} =  \sum_{\mu = 1}^{4}\al^{-2}Q^{\mu}|P_{\mu}|^{(1 - \alpha)}\dfrac{\partial{}}{\partial{P_{\mu}}}\wedge \dfrac{\partial{}}{\partial{Q^{\mu}}} 
\end{equation*}
and $\mathcal{P}_{\al}$ are two compatible $\al$-Poisson bivectors  
with vanishing Schouten-Nijenhuis bracket \\ 
$[\mathcal{P}_{\al},  \mathcal{P}_{\alpha_{1}}]_{NS} = 0.$

Furthermore, we put  $\mathcal{P}^{k +1}_{\alpha_{1}} = S_{k +1} \mathcal{P}_{\alpha_{1}} = S_{k +1} \displaystyle\sum_{\mu = 1}^{4}\al^{-2}Q^{\mu}|P_{\mu}|^{(1 - \alpha)}\dfrac{\partial{}}{\partial{P_{\mu}}}\wedge \dfrac{\partial{}}{\partial{Q^{\mu}}},$ with \\ $ S_{k +1} = \dfrac{1 - k\al }{1 - (k + 1)\al }, k = i+j  \in \mathbb{N}, \ (1 - (k + 1)\al) \neq 0,$
and introduce the following $\al$-Poisson bracket $\{.,.\}^{k_{1}}_{\al_{1}}$
\begin{equation}\label{pbra}
\{f,g\}^{k + 1}_{\al_{1}} := \sum_{\mu = 1}^{4}\al^{-2}S_{k +1}Q^{\mu}|P_{\mu}|^{(1 - \alpha)}\Bigg(\dfrac{\partial{f}}{\partial{P_{\mu}}}\dfrac{\partial{g}}{\partial{Q^{\mu}}} - \dfrac{\partial{f}}{\partial{Q^{\mu}}}\dfrac{\partial{g}}{\partial{P_{\mu}}}\Bigg)
\end{equation}
with respect to the $\al$-symplectic form $  	 \omega^{k + 1}_{\alpha_{1}} =  \displaystyle\sum_{\mu = 1}^{4} \al^{2} S^{-1}_{k + 1}(Q^{\mu})^{-1}|P_{\mu}|^{(\alpha - 1)}dP_{\mu}\wedge dQ^{\mu}$
and get
\begin{equation*}
X_{\al_{k}} = \{H_{\al_{k}},.\}_{\al} = \{H_{\al_{k + 1}},.\}^{k + 1}_{\al_{1}},
\end{equation*}
proving that $X_{\al_{k}}$ are bi-Hamiltonian vector fields defined by the two $\al$-Poisson bivectors $\mathcal{P}_{\al}$ and $\mathcal{P}^{k + 1}_{\al_{1}}.$
Then, the quadruple $(\mathcal{Q},\mathcal{P}_{\al},\mathcal{P}^{k + 1}_{\al_{1}},X_{\al_{k}} )$ is a bi-Hamiltonian system for each $k$.

The associated $\al$-recursion operators are given by
\[
T_{(k+ 1)\al} := \mathcal{P}^{k + 1}_{\al_{1}}\circ \mathcal{P}_{\al}^{-1} = \sum_{\mu = 1}^{4} S_{k} |Q^{\mu}|^{\alpha - 1} Q^{\mu}\bigg( \dfrac{\partial{}}{\partial{P_{\mu}}}\otimes dP_{\mu} + \dfrac{\partial{}}{\partial{Q^{\mu}}}\otimes dQ^{\mu} \bigg).
\]
In addition, we have
{\small\begin{eqnarray*}
		& &\mathcal{L}_{Z_{\al_{0}}} (\mathcal{P}_{\al}) = 0, \ (\tilde{\alpha} = 0), \quad \mathcal{L}_{	Z_{\al_{0}}} (\mathcal{P}^{k + 1}_{\al_{1}}) = - \al \displaystyle\sum_{\mu = 1}^{4}\al^{-2}S_{k +1}Q^{\mu}|P_{\mu}|^{(1 - \alpha)}\dfrac{\partial{}}{\partial{P_{\mu}}}\wedge \dfrac{\partial{}}{\partial{Q^{\mu}}} = - \al\mathcal{P}^{k + 1}_{\al_{1}}, \cr 
		& &  (\tilde{\beta} = -\al), \  \mathcal{L}_{	Z_{\al_{0}}} (H_{\al}) = - Q^{1} = -H_{\al},\ (\tilde{\gamma} = - 1)
\end{eqnarray*}}
permitting to conclude that the vector field
{\small\begin{align*} 
	Z_{\al_{0}}  & = \sum_{\mu = 1}^{4} \bigg( P_{\mu} \dfrac{\partial{}}{\partial{P_{\mu}}} - Q^{\mu}\dfrac{\partial{}}{\partial{Q^{\mu}}}\bigg)
	\end{align*}} 
is a conformal symmetry for  $\mathcal{P}_{\al}, \mathcal{P}^{k + 1}_{\al_{1}}$ and $H_{\al}$ \cite{fer2}.

Defining now the families of quantities   $X^{k + 1}_{\alpha_{l}}, \ Z^{k + 1}_{\al_{l}}, \ \mathcal{	P}^{k + 1}_{\alpha_{l}}, \ \omega^{k + 1}_{\alpha_{l}} $ and $dH^{k + 1}_{\alpha_{l}}$  by

{\small$
	X^{k + 1}_{\alpha_{l}}:=  T^{l}_{(k + 1)\al} X_{\alpha}, \ 	Z^{k + 1}_{\al_{l}}:=  T^{l}_{(k + 1)\al}Z_{\al_{0}}, \ \mathcal{	P}^{k + 1}_{\alpha_{l}}:= T^{l}_{(k + 1)\al}\mathcal{	P}_{\alpha}, \ \omega^{k +1}_{\alpha_{l}}:= ( (T^{l}_{(k + 1)\al})^{\ast})\omega_{\alpha} , \\ $
	$dH^{k +1}_{\alpha_{l}} :=(T^{l}_{(k + 1)\al})^{\ast} dH_{\alpha}$,
}
where $ l \in \mathbb{N}, $ and  $T_{(k + 1)\al}^{\ast} :=  \mathcal{P}_{\alpha}^{-1}\circ \mathcal{	P}^{k +1}_{\alpha_{1}} $ denoting the adjoint of \\ $T_{(k + 1)\al}: = \mathcal{	P}^{k +1}_{\alpha_{1}} \circ  \mathcal{P}_{\alpha}^{-1}, $
we obtain
{\small\begin{align*} 
	X^{k +1}_{\alpha_{l}} & =  -  \al^{-2}(S_{k+1})^{l} (Q^{1})^{1 + \al(l -1)}|P_{1}|^{( 1- \alpha)} \dfrac{\partial{}}{\partial{P_{1}}};\cr
	Z^{k +1}_{\al_{l}} &=  \sum_{\mu = 1}^{4} (S_{k +1})^{l}|Q^{\mu}|^{l(\al - 1)} (Q^{\mu})^{l} \bigg( P_{\mu} \dfrac{\partial{}}{\partial{P_{\mu}}} - Q^{\mu}\dfrac{\partial{}}{\partial{Q^{\mu}}} \bigg);  \\ 
	\mathcal{	P}^{k + 1}_{\alpha_{l}} &= 
	\sum_{\mu = 1}^{4}\al^{-2}(S_{k +1})^{l}(Q^{\mu})^{l}|P_{\mu}|^{(1 -\alpha )}|Q^{\mu}|^{(1 - \alpha)(1 - l)} \dfrac{\partial{}}{\partial{P_{\mu}}}\wedge \dfrac{\partial{}}{\partial{Q^{\mu}}}; \\
	\omega^{k + 1}_{\alpha_{l}} & = 
	\sum_{\mu = 1}^{4} \al^{2}(S_{k +1})^{l}(Q^{\mu})^{l}|P_{\mu}|^{(\alpha - 1 )}|Q^{\mu}|^{(  \alpha - 1)(l + 1)} dP_{\mu}\wedge dQ^{\mu}; \\
	dH^{k +1}_{\alpha_{l}} & = (S_{k +1})^{l}(Q^{1})^{\al l} dQ^{1}; \ \mbox{and} \ H^{k +1}_{\alpha_{l}} = \frac{1}{l\al + 1}(S_{k +1})^{l} (Q^{1})^{\al l + 1}
	\end{align*}}
and for each $l \in \mathbb{N}$ we derive the following plethora of conserved quantities:
{\small\begin{align*}
	\mathcal{L}_{Z^{k +1}_{\al_{l}}}(Z^{k+1}_{\al_{h}}) &  = \al(l - h)(S_{k +1})^{l +h} \sum_{\mu = 1}^{4} |Q^{\mu}|^{(l + h)(\al - 1)} (Q^{\mu})^{l + h} \bigg( P_{\mu} \dfrac{\partial{}}{\partial{P_{\mu}}} - Q^{\mu}\dfrac{\partial{}}{\partial{Q^{\mu}}} \bigg)\cr
	&= \al(l-h)  Z^{k +1}_{\al_{l + h}} ; \cr
	\mathcal{L}_{Z^{k +1}_{\al_{l}}}(X^{k +1}_{\alpha_{h}}) & = \al^{-2} (S_{k + 1})^{l + h}(h\al + 1) (Q^{1})^{1 + \al((l + h) -1)}|P_{1}|^{( 1- \alpha)} \dfrac{\partial{}}{\partial{P_{1}}}\cr
	& = - (h\al + 1)X^{k +1}_{\alpha_{l +h }}   ;
	\end{align*}}
{\small\begin{align*}
	\mathcal{L}_{Z^{k +1}_{\al_{l}}} ( \mathcal{	P}^{k +1}_{\alpha_{h}}) 
	& =  \al^{-1}(S_{k +1})^{l +h}(l - h)(Q^{\mu})^{l+h}|P_{\mu}|^{(1 -\alpha )}|Q^{\mu}|^{(1 - \alpha)(1 - (l +h))} \dfrac{\partial{}}{\partial{P_{\mu}}}\wedge \dfrac{\partial{}}{\partial{Q^{\mu}}} \cr
	& = \al (l - h)\mathcal{P}^{k +1}_{\alpha_{l + h}} ; \\
	\mathcal{L}_{Z^{k +1}_{\al_{l}}} ( \omega^{k +1}_{\alpha_{h}}) & = - \al^{3}(S_{k +1})^{l +h}(l + h)\sum_{\mu = 1}^{4}(Q^{\mu})^{l +h}|P_{\mu}|^{(\alpha - 1 )}|Q^{\mu}|^{(  \alpha - 1)( (l +h) + 1)} dP_{\mu}\wedge dQ^{\mu} \cr &= - \al (l + h) \omega^{k +1}_{\alpha_{ l+ h}};\\
	<dH^{k +1}_{\alpha_{l}}, Z^{k +1}_{\al_{h}} > & = -(S_{k +1})^{l+h}\dfrac{\al (l+h) + 1}{\al (l+h) + 1} (Q^{1})^{  1 + \al (l + h)} \cr
	& = - (\al (l+h) + 1 ) H^{k +1}_{\alpha_{l + h}}; \\
	\mathcal{L}_{Z^{k +1}_{\al_{l}}} (T_{(k +1)\al}) & = -  \al \sum_{\mu = 1}^{4} (S_{k +1})^{l +1}|Q^{\mu}|^{(\alpha - 1)(l +1) }(Q^{\mu})^{l +1}\bigg( \dfrac{\partial{}}{\partial{P_{\mu}}}\otimes dP_{\mu} + \dfrac{\partial{}}{\partial{Q^{\mu}}}\otimes dQ^{\mu} \bigg) \cr
	&= - \al T_{(k + 1)\al}^{l + 1},
	\end{align*}}
satisfying the following relations linking the master symmetries ${Z_{\al_{j}}}$ to the conformal symmetry ${Z_{\al_{0}}}$ for  $\mathcal{P}_{\al}, \mathcal{P}^{k + 1}_{\al_{1}}$ and $H_{\al}$, and to a set of conformal symmetries generated by successive applications of the recursion operator $T_{(k+ 1)\al}$ on  ${Z_{\al_{0}}}$ :
{\small\begin{align*}
	&\mathcal{L}_{Z^{k +1}_{\al_{l}}}(Z^{k +1}_{\al_{h}}) = (\tilde{\beta} - \tilde{\alpha})(h - l)Z^{k +1}_{\al_{l +h}}, \  \mathcal{L}_{Z^{k +1}_{\al_{l}}} (X^{k +1}_{\alpha_{h}}) = (\tilde{\beta} + \tilde{\gamma} + (h - 1)(\tilde{\beta}  -\tilde{\alpha})) X^{k +1}_{\alpha_{l + h}}, \cr &\mathcal{L}_{Z^{k +1}_{\al_{l}}} (\mathcal{	P}^{k +1}_{\alpha_{h}}) = (\tilde{\beta} + (h -l -1)(\tilde{\beta} - \tilde{\alpha})) \mathcal{P}^{k +1}_{\alpha_{l + h}}, \ \mathcal{L}_{Z^{k +1}_{\al_{l}}}(\omega^{k +1}_{\alpha_{h}})=  (\tilde{\beta} + (l + h - 1)(\tilde{\beta} - \tilde{\alpha})) \omega^{k +1}_{\alpha_{ l + h}}, \cr
	& \mathcal{L}_{Z^{k +1}_{\al_{l}}}(T_{(k +1)\al}) = (\tilde{\beta} - \tilde{\alpha}) T_{(k +1)\al}^{1 + l} , \ \langle dH^{k +1}_{\alpha_{h}} , Z^{k +1}_{\al_{l}} \rangle  = 
	(\tilde{\gamma} + (l + h)(\tilde{\beta} - \tilde{\alpha}))H^{k + 1}_{\alpha_{l + h}}. 
	\end{align*}}
This is reminiscent to the well known Oevel formulas (see \cite{oev,fer2,smir,smir2,hkn3,hkn4}).

Finally, it is worth mentioning a  generalization of   the $\al$-Poisson brackets \eqref{pbra} as follows: 
\begin{equation*}
\{f,g\}^{k + t}_{\al_{t}} := \sum_{\mu = 1}^{4}\al^{-2}S_{k +t}|Q^{\mu}|^{1 + \al(t-1)}|P_{\mu}|^{(1 - \alpha)}\Bigg(\dfrac{\partial{f}}{\partial{P_{\mu}}}\dfrac{\partial{g}}{\partial{Q^{\mu}}} - \dfrac{\partial{f}}{\partial{Q^{\mu}}}\dfrac{\partial{g}}{\partial{P_{\mu}}}\Bigg),
\end{equation*}
where $ S_{k +t} = \dfrac{1 - k\al }{1 - (k + t)\al }, k, t  \in \mathbb{N}, \ (1 - (k + t)\al) \neq 0, \ \mbox{and} \  \{f,g\}^{0}_{\al_{0}} = \{f,g\}_{\al}, \  \mbox{with} \ S_{0} = 1,$
leading to a set of $\al-$ generalized bi-Hamiltonian vector fields
\begin{equation*}
X_{\al_{k}} = \{H_{\al_{k}},.\}_{\al} = \{H_{\al_{k + t}},.\}^{k + t}_{\al_{t}},
\end{equation*}
main ingredients governing the Hamiltonian dynamics and pertaining symmetries.

	\section{Concluding remarks} \label{sec6}  
	
In this work, we have proved that a Minkowski phase space endowed with a bracket relatively to a conformable differential realizes a conformable Poisson algebra, confering a bi-Hamiltonian structure to the resulting manifold. We have deduced  that the related $\al$-Hamiltonian vector field for a free particle is an infinitesimal Noether symmetry. We have computed the corresponding $\al-$deformed recursion operator. Using  the Hamiltonian-Jacobi separability, we have constructed  recursion operators in the framework of $\alpha$-Schwarzschild and Friedmann-Lemaître-Robertson-Walker (FLRW) metrics, and obtained related constants of motion.  We have highlighted the existence of a hierarchy of bi-Hamiltonian structures in both the metrics, and derived  a family of $\al$-recursion operators and master symmetries generating the constants of motion. 
This study has also shown that Hamiltonian dynamics hints at a connection between the geometry of our physical system, ($\alpha-$deformes symplectic manifolds and related Hamiltonian vector fields), and conservation laws.
  In this connection, the free particle positions  on the $\alpha$-deformed manifolds are viewed as states and vector fields as laws governing how those states evolve.

	Further, we have calculated the $\al-$deformed Christoffel symbols, Ricci scalar, the components of the $\al-$deformed Riemann, Ricci, and Einstein tensors. This study  revealed  that the $\al$-Christoffel symbols ($	(\Gamma^{1}_{11})_{\alpha}, 	(\Gamma^{2}_{22})_{\alpha}, 	(\Gamma^{3}_{33})_{\alpha}, \ \mbox{and}\	(\Gamma^{4}_{44})_{\alpha}$) in  $\al-$deformed Minkowski metric are no longer null, contrary to the undeformed case corresponding to ${\alpha}=1$. Similarly, the  $\al$-Christoffel symbols  ($	(\Gamma^{1}_{11})_{\alpha}, 	(\Gamma^{3}_{33})_{\alpha}, \ \mbox{and}\	(\Gamma^{4}_{44})_{\alpha}$) are not equal zero in  $\al-$deformed Schwarzschild and FLRW metrics. The existence of these  symbols $(\Gamma^{i}_{ii})_{\alpha}, (i = 1,2,3,4)$  informs us about the way in which the parallel displacement of any basic vector on the $\al$-deformed manifolds with respect to itself always remains parallel to the same basic vector.
	\subsection*{Acknowledgments}	
	The ICMPA-UNESCO Chair is in partnership
	with the Association pour la Promotion Scientifique de l'Afrique
	(APSA), France, and Daniel Iagolnitzer Foundation (DIF), France,
	supporting the development of mathematical physics in Africa.
	M. M. is supported by the Faculty of Mechanical Engineering,
	University of Ni\v s, Serbia, Grant ``Research and development of
	new generation machine systems in the function of the technological
	development of Serbia''.
	
\appendix

\section{Appendix}
\begin{landscape}
	\begin{center}
		\subsection{$\al$-Christoffel symbols $(\Gamma^{k}_{ij})_{\alpha}$ in $\al$-Schwarzschild metric} 
		\begin{tabular}{|p{5 cm}|p{7 cm}|p{7 cm}|p{4cm}|}
			\hline
			$(\Gamma^{1}_{11})_{\alpha} = \dfrac{\alpha - 1}{q^{1}}$ & $(\Gamma^{2}_{11})_{\alpha} = - \dfrac{M(2M -q^{2})(q^{1})^{2(\al - 1)}}{(q^{2})^{2\al + 1}}$ &  &  \\ 
			\hline
			$(\Gamma^{1}_{12})_{\alpha} = - \dfrac{M}{q^{2}(2M -q^{2})}$ & $(\Gamma^{2}_{22})_{\alpha} = \dfrac{(1 - \al)q^{2} + (2\al -1)M}{q^{2}(2M -q^{2})}$ & $(\Gamma^{3}_{23})_{\alpha} = \dfrac{1}{q^{2}}$ & $(\Gamma^{4}_{24})_{\alpha} = \dfrac{1}{q^{2}}$ \\
			\hline
			& $(\Gamma^{2}_{33})_{\alpha} = - \dfrac{(2M -q^{2})(q^{3})^{2(\al - 1)}}{(q^{2})^{2(\al - 1)}}$ & $(\Gamma^{3}_{33})_{\alpha} = \dfrac{\al - 1}{q^{3}}$ & $(\Gamma^{4}_{34})_{\alpha} = \cot(q^{3})$ \\
			\hline
			& $(\Gamma^{2}_{44})_{\alpha} = \dfrac{(2M -q^{2})(q^{4})^{2(\al - 1)}\sin^{2}(q^{3})}{(q^{2})^{2(\al - 1)}}$ & $(\Gamma^{3}_{44})_{\alpha} = -\dfrac{(q^{4})^{2(\al - 1)}\sin(q^{3})\cos(q^{3})}{(q^{3})^{2(\al - 1)}}$ & $(\Gamma^{4}_{44})_{\alpha} = \dfrac{\al - 1}{q^{4}}$ \\
			\hline
		\end{tabular}
	\end{center}
\hspace{60pt}
	\begin{center}
		\subsection{ Components of the $\al$-Riemann tensor $\al$-$(R_{ijkl})_{\alpha}$ in $\al$-Schwarzschild metric} 
		\begin{tabular}{|p{6 cm}|p{8 cm}|p{9.5 cm}|}
			\hline
			{\[(R_{1212})_{\alpha} = - \dfrac{(1 + \al)\al^{2}M(q^{1})^{2(\al - 1)} }{(q^{2})^{3}}\]} & {\[(R_{1313})_{\alpha} = -\dfrac{\al^{2}M(q^{1})^{2(\al - 1)}(q^{3})^{2(\al - 1)} (2M -q^{2})  }{(q^{2})^{2\al}}\]} & {\[(R_{1414})_{\alpha} = - \dfrac{\al^{2}M(q^{1})^{2(\al - 1)}(q^{4})^{2(\al - 1)} \sin^{2}(q^{3})(2M -q^{2})  }{(q^{2})^{2\al}}\]} \\
			\hline
			{\begin{align*}
				(R_{2323})_{\alpha} & = \dfrac{\al^{2}(M(2\al - 1)}{2M -q^{2}} \\
				& \ + \dfrac{(1 - \al)q^{2})(q^{3})^{2(\al - 1)}  }{2M -q^{2}}
				\end{align*}} & {\begin{align*}
				(R_{2424})_{\alpha} & = \dfrac{\al^{2}(M(2\al - 1)}{2M -q^{2}} \\
				& \ + \dfrac{(1 - \al)q^{2})(q^{4})^{2(\al - 1)}\sin^{2}(q^{3})  }{2M -q^{2}}
				\end{align*}} & {\begin{align*} 
				(R_{3434})_{\alpha} & = -\dfrac{\al^{2}q^{2}(q^{4})^{2(\al - 1)}}{(q^{2})^{2(\al - 1)}q^{3}} \bigg[\sin^{2}(q^{3}) ( q^{2}q^{3}((q^{3})^{2(\al - 1)} \\
				& \ - (q^{2})^{2(\al - 1)})   - 2M (q^{3})^{2\al - 1} ) \\
				& \ + (1-\al)(q^{2})^{2(\al - 1)}\sin(q^{3})\cos(q^{3})\bigg]
				\end{align*}} \\
			\hline
		\end{tabular}
	\end{center}
\end{landscape}
\begin{landscape}
		\begin{center}
			\subsection{ Components of the $\al$-Ricci tensor $(R_{ij})_{\alpha}$ in $\al$-Schwarzschild metric} 
			\begin{tabular}{|p{10 cm}|p{10 cm}|}
				\hline
				{\[(R_{11})_{\alpha} = - \dfrac{(\al - 1)M(2M -q^{2})(q^{1})^{2(\al - 1)}}{(q^{2})^{2(\al - 1)}}\]} & {\[(R_{22})_{\alpha} = -\dfrac{(1-\al)(3M -2q^{2})}{(q^{2})^{2}(-2M + q^{2})}\]} \\
				\hline
				{\begin{align*} 
					(R_{33})_{\alpha} & = \dfrac{1}{(q^{2})^{2(\al - 1)}q^{3}\sin(q^{3})} \bigg[(1 - \al)(q^{2})^{2(\al - 1)}\cos(q^{3}) \\
					& \ + q^{2}q^{3}[(2 - \al)(q^{3})^{2(\al - 1)}  - (q^{2})^{2(\al - 1)} ]\sin(q^{3}) \\
					& \ + 2(\al - 1)M(q^{2})^{2\al - 1}\sin(q^{3})\bigg]
					\end{align*}} & {\begin{align*} 
					(R_{44})_{\alpha} & = -\dfrac{1}{(q^{2})^{2\al - 1}(q^{3})^{2\al - 1}}\bigg[(q^{4})^{2(\al - 1)}[(\al - 2)q^{2}(q^{3})^{2\al - 1} \\
					& \ + (q^{2})^{2\al - 1}q^{3} + 2(1 - \al)M(q^{3})^{2\al - 1}] \sin^{2}(q^{3}) \\
					& \ + (\al -1)(q^{2})^{2\al - 1}\sin(q^{3})\cos(q^{3})\bigg]
					\end{align*}} \\
				\hline 
			\end{tabular}
		\end{center}
	\hspace{60pt} 
	\begin{center}
		\subsection{$\al$-Ricci scalar $\mathbf{R}$ in $\al$-Schwarzschild metric} 
		\begin{tabular}{|p{20cm}|}
			\hline
			\begin{align*} 
			\mathbf{R} = \dfrac{2[(1 - \al)(q^{2})^{2\al - 1}\cos(q^{3}) + (3(\al -1)M(q^{3})^{2\al - 1} + (3 - 2\al)q^{2}(q^{3})^{2\al - 1} - q^{3}(q^{2})^{2\al - 1})\sin(q^{3})]}{\al^{2}(q^{2})^{2\al + 1}(q^{3})^{2\al - 1}}
			\end{align*} \\	
			\hline
		\end{tabular}
	\end{center}
\end{landscape}
\begin{landscape}
	\begin{center}
			\subsection{Components of the $\al$-Einstein tensor $G_{ij}$ in $\al$-Schwarzschild metric} 	
		\begin{tabular}{|p{7 cm}|p{7.5 cm}|}
			\hline
			{\begin{align*} 
				(G_{11})_{\alpha} & = - \dfrac{1}{(q^{1})^{2(1 - \al )}(q^{2})^{2(\al + 1)}(q^{3})^{2\al - 1}}\bigg[(2M -q^{2})[(1 - \al)(q^{2})^{2\al - 1}\cos(q^{3}) \\
				& \ + (4(\al -1)M(q^{3})^{2\al - 1} + (3 - 2\al)q^{2}(q^{3})^{2\al - 1} - q^{3}(q^{2})^{2\al - 1})\sin(q^{3})]\bigg]
				\end{align*}} & {\begin{align*} 
				(G_{22})_{\alpha} & = - \dfrac{q^{3}\bigg((q^{2})^{2 (\al - 1)} - (q^{3})^{2 (\al - 1)}\bigg)}{q^{2}(2M -q^{2})(q^{3})^{2\al - 1}} \\
				& \ - \dfrac{(\al - 1)(q^{2})^{2 (\al - 1)}\cot(q^{3})}{q^{2}(2M -q^{2})(q^{3})^{2\al - 1}}
				\end{align*}} \\
			\hline
			{\[(G_{33})_{\alpha} = \dfrac{(\al - 1)(q^{2} - M)(q^{3})^{2(\al - 1)}}{(q^{2})^{2\al - 1}}\]} & {\[(G_{44})_{\alpha} = \dfrac{(\al - 1)(q^{2} - M)(q^{4})^{2(\al - 1)}\sin^{2}(q^{3})}{(q^{2})^{2\al - 1}}\]} \\
			\hline
		\end{tabular}
	\end{center}
\end{landscape}
\begin{landscape}
	\begin{center}
		\subsection{$\al$-Christoffel symbols $(\Gamma^{k}_{ij})_{\alpha}$ in $\al$-FLRW metric}
		\begin{tabular}{|p{6.2 cm}|p{6 cm}|p{3 cm}|p{4.8 cm}|}
			\hline
			{\[(\Gamma^{1}_{11})_{\alpha} =  {\displaystyle \frac {\alpha  - 1}{q^{1}}}\]} & {\[(\Gamma^{2}_{12})_{\alpha} =   \dfrac {1}{2} \,
				{\displaystyle \frac {{\dfrac {d R(q^{1})}{dq^{1}}}\,
					}{ R(q^{1})} }\]} & {\[(\Gamma^{3}_{13})_{\alpha}= \dfrac {1}{2} \,
				{\displaystyle \frac {{\dfrac {dR(q^{1})}{dq^{1}}}\,
					}{ R(q^{1})} }\]} & {\[(\Gamma^{4}_{14})_{\alpha} =\dfrac {1}{2} \,
				{\displaystyle \frac {{\dfrac {dR(q^{1})}{dq^{1}}}}{ R(q^{1})} }\]} \\
			\hline
			{\[(\Gamma^{1}_{22})_{\alpha}= - {\displaystyle \frac {1}{2}} \,{\displaystyle 
					\dfrac {{\dfrac {dR(q^{1})}{dq^{1}}}\,(q^{2})^{(2\,\alpha  - 2)}}{ (- 1 + k\,
						(q^{2})^{2})(q^{1})^{(2\,\alpha  - 2)} }}\]} & {\begin{align*}
				(\Gamma^{2}_{22})_{\alpha} & = - {\displaystyle \dfrac {\alpha ( - 1
						+  \,k\,(q^{2})^{2})}{
						q^{2}\,( - 1 + k\,(q^{2})^{2})}} \\
				& \ - {\displaystyle \dfrac {1 - 2\,k\,(q^{2})^{2}}{
						q^{2}\,( - 1 + k\,(q^{2})^{2})}}
				\end{align*}} & {\[(\Gamma^{3}_{23})_{\alpha}=\dfrac{1}{q^{2}}\]} & {\[(\Gamma^{4}_{24})_{\alpha}=\dfrac{1}{q^{2}}\]} \\
			\hline
			{\begin{align*}
				(\Gamma^{1}_{33})_{\alpha} & =  {\displaystyle \dfrac {1}{2(q^{1})^{(2\,\alpha  - 2)}}} \,
				{\dfrac {dR(q^{1})}{dq^{1}}}\times\, \\
				& \ (q^{2})^{2}\,(q^{3})^{(2\,\alpha  - 2)}
				\end{align*}} & {\[(\Gamma^{2}_{33})_{\alpha}= \frac{q^{2}\,(q^{3})^{(2\,\alpha  - 2)} ( - 1 + k\,(q^{2})^{2})}{(q^{2})^{(2\,\alpha  - 2)}}\]} & {\[(\Gamma^{3}_{33})_{\alpha}={\displaystyle \dfrac {\alpha  - 1}{q^{3}}}\]} & {\[(\Gamma^{4}_{34})_{\alpha}=  \cot(q^{3})\]} \\
			\hline
			{\begin{align*}
				(\Gamma^{1}_{44})_{\alpha} & = {\displaystyle \dfrac {1}{2(q^{1})^{(2\,\alpha  - 2)}}} \,
				{\dfrac {dR(q^{1})}{dq^{1}}}\,\times \\
				& \ (q^{2})^{2}\,(q^{4})^{(2\,\alpha  - 2)}\,\sin^{2}
				(q^{3})
				\end{align*}} & {\begin{align*}
				(\Gamma^{2}_{44})_{\alpha} & = 
				\dfrac{q^{2}\,(q^{4})^{(2\,\alpha  - 2)}\,\sin^{2}(
					q^{3})}{(q^{2})^{(2\,\alpha  - 2)}}\times \\
				& \ ( - 1 + k\,(q^{2})^{2})
				\end{align*}} & {\begin{align*}
				(\Gamma^{3}_{44})_{\alpha} & = - 
				\dfrac{(q^{4})^{(2\,\alpha  - 2)}\,}{(q^{3})^{(2\,\alpha  - 2)}} \times \\
				& \sin(q^{3})\,\cos(q^{3})
				\end{align*}} & {\[(\Gamma^{4}_{44})_{\alpha}={\displaystyle \frac {\alpha  - 1}{q^{4}}}\]} \\
			\hline
		\end{tabular}
	\end{center}
\end{landscape}
\begin{landscape}
	\begin{center}
			\subsection{Components of the $\al$-Riemann tensor $(R_{ijkl})_{\alpha}$  in $\al$-FLRW metric}
		\begin{tabular}{|p{0.5 cm}|p{0.5cm}|p{0.5cm}|}
			\hline
			{\begin{align*}
				R_{1212} & = - \dfrac{ \alpha ^{2}\,
					(q^{2})^{(2\,\alpha  - 2)}}{4( - 1 + k\,
					(q^{2})^{2})\,q^{1}\,R(q^{1})}\times \\
				& \ \bigg[ - 2\,\dfrac {d^{2}R(q^{1})}{(dq^{1})^{2}}\,q^{1}\,R(q^{1}) \\
				& \ + 2\,\dfrac {dR(q^{1})}{dq^{1}}\,R(q^{1})(\alpha -1) \\
				& \ + \bigg(\dfrac {d\,R(q^{1})}{dq^{1}}\bigg)^{2}\,q^{1}\bigg]
				\end{align*}} & {\small\begin{align*}
				R_{1313} & ={\displaystyle \dfrac {\alpha ^{2}\,
						(q^{2})^{2}\,(q^{3})^{(2\,\alpha  - 2)}}{4q^{1}\,
						R(q^{1})}} \times \\
				& \ \bigg[ - 2\,\dfrac {d^{2}R(q^{1})}{(dq^{1})^{2}}\,q^{1}\,R(q^{1}) \\ 
				& \ + 2\,\dfrac {dR(q^{1})}{dq^{1}}\,R(q^{1})(\alpha -1) \\
				& \ + \bigg(\dfrac {d\,R(q^{1})}{dq^{1}}\bigg)^{2}\,q^{1}\bigg]
				\end{align*}} & {\small\begin{align*}
				R_{1414} & ={\displaystyle \dfrac {\alpha ^{2}\,
						(q^{2})^{2}\,(q^{4})^{(2\,\alpha  - 2)}\,\sin
						(q^{3})^{2}}{4q^{1} 
						R(q^{1})}}\times \\
				& \ \bigg[ - 2\,\dfrac {d^{2}R(q^{1})}{(dq^{1})^{2}}\,q^{1}\,R(q^{1}) + 2\,\dfrac {dR(q^{1})}{dq^{1}}\,R(q^{1})(\alpha -1) \\
				& \ + \bigg(\dfrac {dR(q^{1})}{dq^{1}}\bigg)^{2}\,q^{1}\bigg]
				\end{align*}} \\
			\hline
			{\small\begin{align*}
				R_{2323}&={\displaystyle \frac {\alpha ^{2}\,
						(q^{3})^{(2\,\alpha  - 2)}}{4\bigg(
						(q^{1})^{(2\,\alpha  - 2)}\,( - 1 + k\,(q^{2})^{2})\bigg)}} \times \\
				& \ \bigg[- \bigg({\dfrac {dR(q^{1})}{dq^{1}}}\,\bigg)^{2}\,(q^{2})^{2}\,(q^{2})^{(2\,\alpha  - 2)} \\
				& \ + 4\alpha\,R(q^{1})\,(q^{1})^{(2\,\alpha (  - 2))}\, (-1 + k(q^{2})^{2}) \\
				& \ + 4\, R(q^{1})\,(q^{1})^{(2\,\alpha  - 2)}(  1 - 2k\,(q^{2})^{2}) \bigg]
				\end{align*}} & {\small\begin{align*}
				R_{2424}&={\displaystyle \frac {\alpha ^{2}\,
						(q^{4})^{(2\,\alpha  - 2)}\,\sin(q^{3})^{2}}{4(
						(q^{1})^{(2\,\alpha  - 2)}\,( - 1 + k\,(q^{2})^{2}))}} \times \\
				& \ \bigg[- \bigg({\dfrac {d}{dq^{1}}}
				\,R(q^{1})\bigg)^{2}\,(q^{2})^{2}\,(q^{2})^{(2\,\alpha  - 2)} \\
				& \ + 4\alpha\,R(q^{1})\,(q^{1})^{(2\,\alpha  - 2)}\, (-1 + k(q^{2})^{2}) \\
				& \ + 4\, R(q^{1})\,(q^{1})^{(2\,\alpha  - 2)}(  1 - 2k\,(q^{2})^{2}) \bigg]
				\end{align*}} & {\small\begin{align*}
				R_{3434}& ={\displaystyle \frac {\alpha ^{2}\,
						(q^{2})^{2}\,(q^{4})^{(2\,\alpha  - 2)}\sin^{2}(q^{3})}{4(q^{1})^{(2\,\alpha 
							- 2)}\,(q^{2})^{(2\,\alpha  - 2)}\,q^{3}}} \times \\
				& \ \bigg[  4\,
				R(q^{1})\,
				(q^{1})^{(2\,\alpha  - 2)}\bigg((q^{2})^{(2\,\alpha  - 2)} - (q^{3})^{(2\,\alpha  - 2)}\bigg)q^{3} \\
				& \ + \bigg({\frac {dR(q^{1})}{dq^{1}}}\bigg)^{2}\,(q^{2})^{2}\,(q^{3})^{(2\,\alpha  - 2)}\,
				(q^{2})^{(2\,\alpha  - 2)}\,q^{3} \\
				& \ +  4\,R(q^{1})(q^{1})^{(2\,\alpha  - 2)}(q^{3})^{(2\,\alpha
					- 2)}\,q^{3}\,k\,
				(q^{2})^{2} \\
				& + 4\,(\alpha - 1)\,R(q^{1})\,(q^{1})^{(2\,\alpha  - 2)}
				\,(q^{2})^{(2\,\alpha  - 2)} \cot(q^{3}) \bigg]
				\end{align*}} \\
			\hline
		\end{tabular}
	\end{center}
\end{landscape}
\begin{landscape}
	\begin{center}
			\subsection{ Components of the $\al$-Ricci tensor $(R_{ii})_{\alpha}$ in $\al$-FLRW metric}
		\begin{tabular}{|p{9cm}|p{9 cm}|}
			\hline
			{\begin{align*}
				R_{11}&= - {\displaystyle \dfrac {3}{4R(q^{1})^{2}\,q^{1}}} \bigg[ - 2\,{\frac {d^{2}R(q^{1})}{(dq^{1})^{2}}}\,q^{1}\,R(q^{1}) \\
				& \ + 2\,{\frac {dR(q^{1})}{dq^{1}
				}}\,R(q^{1})\,(\alpha -1)
				+ \bigg({\frac {dR(q^{1})}{dq^{1}}}\bigg)^{2}\, q^{1} \bigg]
				\end{align*}} & {\begin{align*}
				R_{22}&= - {\displaystyle \frac {1}{4(q^{1})^{(
							2\,\alpha  - 2)}\,( - 1 + k\,(q^{2})^{2})\,q^{1}\,
						R(q^{1})\,(q^{2})^{2} }} \times \\
				& \ \bigg[ - 2\,(q^{2})^{(2\,\alpha  - 2)}\,(q^{2})^{2}\,{\frac {d^{2}R(q^{1})}{(d
						q^{1})^{2}}}\,q^{1}\,
				R(q^{1}) \\
				& \ + 2\,(q^{2})^{(2\,\alpha  - 2)}\,(q^{2})^{2}
				\,{\frac {dR(q^{1})}{dq^{1}}}\,
				R(q^{1})\,(\alpha - 1) \\
				& \ - (q^{2})^{(2\,\alpha  - 2)}\,(q^{2})^{2}\,\bigg(
				{\dfrac {dR(q^{1})}{dq^{1}}} \bigg)^{2}\,
				q^{1} \\
				& \ + 8\,q^{1}\,R(q^{1})\,
				(q^{1})^{(2\,\alpha  - 2)}\,\alpha (-1  + k\,(q^{2})^{2}) \\
				& \ + 8\,
				q^{1}\,R(q^{1})\,(q^{1})^{(2\,
					\alpha  - 2)} (1 - 2k\,(q^{2})^{2})\bigg]
				\end{align*}} \\
			\hline
			{\begin{align*}
				R_{33}&={\displaystyle \dfrac {1}{4 
						\,q^{1}\,R(q^{1})\,(q^{1})^{(2\,\alpha  - 2)}\,(q^{2})^{(2\,
							\alpha  - 2)}\,q^{3}}} \times \\
				& \ \bigg[4\,q^{1}\,
				R(q^{1})\,(q^{1})^{(2\,\alpha  - 2)}\,(q^{2})^{(2\,\alpha  - 2)}\bigg(\cot(q^{3})(1-\alpha) -  q^{3}\bigg)\\
				& \ + 4\,q^{1}\,R(q^{1})\,(q^{1})^{(2\,\alpha  - 2)}\,q^{3} (q^{3})^{(2\,\alpha  - 2)} \times \\ 
				& \ \bigg(2 - 3k\,(q^{2})^{2} + \alpha (-1 + k\,(q^{2})^{2})  \bigg) \\
				& \ - 2\,(q^{2})^{2}\,
				(q^{3})^{(2\,\alpha  - 2)}\,(q^{2})^{(2\,\alpha  - 2)
				}\,q^{3}\, {\frac {d^{2}R(q^{1})}{(dq^{1})^{2}}}\,q^{1}\,R(q^{1}) \\
				& \ + {\frac {dR(q^{1})
					}{dq^{1}}} \,(q^{2})
				^{2}\,(q^{3})^{(2\,\alpha  - 2)}\,(q^{2})^{(2\,\alpha
					- 2)}\,q^{3} \times \\
				& \ \bigg(-\,q^{1}\,{\frac {dR(q^{1})
					}{dq^{1}}} + 2(\alpha -1)R(q^{1}) \bigg)  \bigg]
				\end{align*}} & {\begin{align*}
				R_{44} &= - {\displaystyle \dfrac {(q^{4})^{(2
							\,\alpha  - 2)}\sin^{2}(q^{3})}{4q^{1}\,R(q^{1})\,(q^{1})^{(2\,\alpha  - 2)
						} (q^{2})^{(2\,\alpha  - 2)}\,(q^{3})^{(2\,\alpha  - 2)
						}\,q^{3}}} \times \\
				& \ \bigg[2\,(q^{2})^{2}\,(q^{3})^{(2\,\alpha  - 2)}
				\,(q^{2})^{(2\,\alpha  - 2)}\,q^{3}
				\,R(q^{1}) \times \\
				& \ \bigg( {\dfrac {d^{2}(R(q^{1}))
					}{d(q^{1})^{2}}}\,q^{1} - (\al - 1){\frac {d(R(q^{1}))}{dq^{1}}} \bigg)    \\
				& \ + 4\,(q^{3})^{(2\,\alpha  - 2)}\,q^{1}\,
				q^{3}\,R(q^{1})\,(q^{1})^{(2\,
					\alpha  - 2)}\bigg(\al(1 - k\,(q^{2})^{2}) -2 + 3k\,(q^{2})^{2}\bigg) \\
				& \ + 4\,q^{1}\,R(q^{1})\,(q^{1})^{(2\,\alpha  - 2)}\,
				(q^{2})^{(2\,\alpha  - 2)} \bigg(q^{3} + (\al -1)\cot(q^{3})  \bigg) \\
				& \ + q^{1}\,\bigg({\frac {d R
						(q^{1})}{dq^{1}}}\bigg)^{2}\,(q^{2})^{2}\,(q^{3})^{(2\,
					\alpha  - 2)}\,(q^{2})^{(2\,\alpha  - 2)}\,q^{3}  \bigg]
				\end{align*}} \\
			\hline
		\end{tabular}
	\end{center}
\end{landscape}
\begin{landscape}
	\begin{center}
			\subsection{ $\al$-Ricci scalar $\mathbf{R}$ in $\al$-FLRW metric}
		\begin{tabular}{|p{22cm}|}
			\hline
			{\small\begin{align*} 
				\mathbf{R}&= \dfrac{1}{(R(q^{1})\,(q^{2})^{2} 
					\alpha ^{2}\,q^{1}\,(q^{1})^{(2\,\alpha  - 2)}\,
					(q^{2})^{(2\,\alpha  - 2)}\,(q^{3})^{(2\,\alpha  - 2)
					}\,q^{3})}\bigg[2\,(1-\alpha)q^{1}\,\cot(q^{3})\,
				(q^{1})^{(2\,\alpha  - 2)}\,(q^{2})^{(2\,\alpha  - 2)
				} \cr
				& + 3(\alpha -1)\,(q^{2})^{2}\,
				(q^{3})^{(2\,\alpha  - 2)}\,(q^{2})^{(2\,\alpha  - 2)
				}\,q^{3}\,{\dfrac {dR(q^{1})}{dq^{1}}} \cr
				& + 4\alpha\,(q^{3})^{(2\,
					\alpha  - 2)}\,q^{1}\,q^{3}\,(q^{1})^{(2\,
					\alpha  - 2)}\, (-1 + k\,(q^{2})^{2}) + 2\,q^{1}\,(q^{1})^{(2\,\alpha  - 2)}\,
				q^{3} \bigg(3 (q^{3})^{(2\,\alpha  - 2)} - (q^{2})^{(2\,\alpha  - 2)}\bigg) \cr
				&- 3\,(q^{2})^{2}\,
				(q^{3})^{(2\,\alpha  - 2)}\,(q^{2})^{(2\,\alpha  - 2)
				}\,q^{3}\,{\frac {d^{2}R
					(q^{1})}{(dq^{1})^{2}}}\,q^{1} - 10\, q^{1}\,
				(q^{3})^{(2\,\alpha  - 2)}\,(q^{1})^{(2\,\alpha  - 2)
				}\,q^{3}\,k\,(q^{2})^{2},   \bigg] 
				\end{align*}} \\			
			\hline
		\end{tabular}
	\end{center}
\end{landscape}			
\begin{landscape}
	\begin{center}
			\subsection{ Components of the $\al$-Eintein tensor $(G_{ij})_{\alpha}$ in $\al$-FLRW metric}
		\begin{tabular}{|p{10 cm}|p{10 cm}|}
			\hline
			{\begin{align*}
				G_{11} &= - {\displaystyle \frac {1}{4(
						R(q^{1})^{2}\,
						(q^{2})^{2}\,(q^{2})^{(2\,\alpha  - 2)}\,(q^{3})^{(2\,\alpha  - 2)}\,q^{3}) }} \times \\
				& \ \bigg[3\,\,\bigg({\frac {dR(q^{1})}{dq^{1}}}\,\bigg)^{2}\,(q^{2})^{2}\,(q^{3})^{(2\,\alpha  - 2)}
				\,(q^{2})^{(2\,\alpha  - 2)}\,q^{3} \\
				& + 4(\alpha - 1)\,R(q^{1})\,\cot(q^{3})\,(q^{1})^{(2\,\alpha  - 2)}\,(q^{2})^{(2\,\alpha 
					- 2)} \\
				& \ + 4\,R(q^{1})\,(q^{1})^{(2\,\alpha  - 2)}\,(q^{2})^{(2\,\alpha 
					- 2)}\,q^{3}\cr
				& + \ 4\,R(q^{1})\,(q^{3})^{(2\,\alpha  - 2)}\,(q^{1})^{(2\,
					\alpha  - 2)}\,q^{3} \bigg( (-3 + 5\,k\,(q^{2})^{2}) \\
				& \ - 2\al (-1 + k\,(q^{2})^{2}) \bigg)   \bigg]
				\end{align*}} & {\begin{align*}
				G_{22}& ={\displaystyle \frac {1}{4((q^{1})^{(2\,\alpha  - 2)}\,( - 1 + k\,(q^{2})^{2}) 
						q^{1}\,R(q^{1})\,(q^{2})^{2}\,(q^{3})^{(2\,\alpha  - 2)}\,
						q^{3}) }} \\
				& \ \bigg[  - 4\,(q^{2})^{2}\,(q^{3})^{(2\,\alpha  - 2
					)}\,(q^{2})^{(2\,\alpha  - 2)}\,q^{3}\,{\frac {d^{
							2}R(q^{1})}{(dq^{1})^{2}}}\,q^{1}\,R(q^{1}) \\
				& + (q^{2})^{2}\,
				(q^{3})^{(2\,\alpha  - 2)}\,(q^{2})^{(2\,\alpha  - 2)
				}\,q^{3}\,{\frac {dR(q^{1})}{dq^{1}}}\times \\
				& \ \bigg( 4(\alpha - 1)\,R(q^{1})  + q^{1}{\frac {dR(q^{1})
					}{dq^{1}}}\bigg)\\
				& + 4\,q^{1}\,
				R(q^{1})(q^{1})^{(2\,\alpha  - 2)} \bigg(-(q^{3})^{(2\,\alpha  - 2)}\,q^{3}(-1 + k\,(q^{2})^{2}) \\
				& \  + (q^{2})^{(2\,\alpha  - 2)}( - q^{3} + (1-\al)\cot(q^{3}))\bigg) \bigg]
				\end{align*}} \\
			\hline
			{\begin{align*}
				G_{33} &= - {\displaystyle \frac {(q^{3})^{(2
							\,\alpha  - 2)}}{4
						q^{1}\,R(q^{1})\,(q^{1})^{(2\,
							\alpha  - 2)}\,(q^{2})^{(2\,\alpha  - 2)}}}\times \\
				& \ \bigg[4(1-\al)\,q^{1}\,R(q^{1})\,
				(q^{1})^{(2\,\alpha  - 2)} \\
				& \ + 4(\al - 1)\,(q^{2})^{(2\,\alpha 
					- 2)}\,(q^{2})^{2}\,{\frac {d R(q^{1})}{dq^{1}}}\,R(q^{1}) \\
				& \ - 4\,(q^{2})^{(2\,\alpha  - 2)}\,(q^{2})^{2}
				\,{\frac {d^{2}R(q^{1})}{(dq^{1})^{2}}}
				\,q^{1}\,R(q^{1})  \\
				& \ - 4(2-\al)\,q^{1}\,
				R(q^{1})\,(q^{1})^{(2\,\alpha  - 2)}\,k\,
				(q^{2})^{2} \\
				& \ + (q^{2})^{(2\,\alpha  - 2)}\,(q^{2})^{2}\,\bigg(
				{\frac {dR(q^{1})}{dq^{1}}}\bigg)^{2}\,
				q^{1}  \bigg]
				\end{align*}} & {\begin{align*}
				G_{44}&={\displaystyle \frac {(q^{4})^{(2\,
							\alpha  - 2)}\sin^{2}(q^{3})}{4q^{1}\,R(q^{1})\,(q^{1})^{(2\,\alpha  - 2)}\,(q^{2})^{(2\,\alpha 
							- 2)}}} \times \\
				& \ \bigg[  4(\al - 1)\,q^{1}\,R(q^{1})\,
				(q^{1})^{(2\,\alpha  - 2)} \\
				& \ + 4\,(q^{2})^{(2\,\alpha  - 2)}\,(q^{2})^{2}
				\,{\frac {d^{2}R(q^{1})}{(dq^{1})^{2}}}q^{1}\,R(q^{1}) \\
				& \ - 4(\al - 1)\,(q^{2})^{(2\,\alpha  - 2)}\,(q^{2})^{2}
				\,{\frac {dR(q^{1})}{dq^{1}}}\,
				R(q^{1}) \\
				&  + 4\,
				q^{1}\,R(q^{1})\,(q^{1})^{(2\,
					\alpha  - 2)}\,k\,(q^{2})^{2} (2-\al) \\
				& \ - q^{1}\,\bigg({\frac {dR(q^{1})}{dq^{1}}}\bigg)^{2}\,(q^{2})^{2}\,(q^{2})^{(2\,
					\alpha  - 2)}\bigg]
				\end{align*}} \\
			\hline
		\end{tabular}
	\end{center}
\end{landscape}
	
\end{document}